\documentclass[preprint,superscriptaddress,showpacs]{revtex4-1}
\usepackage{lmodern}%modern Latin fonts: Roman, Sans Serif, Typewriter
\usepackage{amsmath,amssymb,amsthm} %math tool
\usepackage{grffile} %graphic tool <--allows blanks in the file name
\usepackage{sidecap} %side caption tool
\usepackage{enumitem} %list/enumerate item tool
\usepackage{array}
\usepackage{multirow}
\usepackage{ctable} %\specialrule command for tables
\usepackage{subfigure}	%use subfigure
\usepackage{hyperref}

\newcommand{\GeV}{$\mathrm{GeV}/{c^{2}}$}

\newcommand{\SK} {$B^+ \to \bar{p} \Lambda K^+ K^+$}
\newcommand{\OK} {$B^+ \to p \bar{\Lambda} K^+ K^-$}
\newcommand{\ETAC} {$B^+ \rightarrow \eta_c K^+$}
\newcommand{\JPSI} {$B^+ \rightarrow J/\psi K^+$}
\newcommand{\CHIC} {$B^+ \rightarrow \chi_{c1} K^+$}
\newcommand{\PLPHI} {$B^+ \rightarrow p \bar{\Lambda} \phi$}
\newcommand{\SKPHSP} {$B^+ \rightarrow \bar{p} \Lambda K^+ K^+$}
\newcommand{\OKPHSP} {$B^+ \rightarrow p \bar{\Lambda} K^+ K^-$}
\newcommand{\LLKO} {$B^+ \rightarrow \Lambda(1520) \bar{\Lambda} K^+$}
\newcommand{\LLKS} {$B^+ \rightarrow \bar{\Lambda}(1520) \Lambda K^+$}
\newcommand{\ETACS} {$\eta_c \rightarrow \Lambda(1520) \bar{\Lambda}$}
\newcommand{\JPSIS} {$J/\psi \rightarrow \Lambda(1520) \bar{\Lambda}$}
\newcommand{\ETACD} {$\eta_c \rightarrow p\bar{\Lambda}K^-$}
\newcommand{\JPSID} {$J/\psi \rightarrow p\bar{\Lambda}K^-$}
\newcommand{\CHICD} {$\chi_{c1} \rightarrow p\bar{\Lambda}K^-$}

\begin{document}

%\preprint{preprint submitted to Phys. Rev. Lett.}
\preprint{\vbox{   \hbox{Belle Preprint 2018-15}
                   \hbox{KEK Preprint 2018-21}
}}

\title{Observation of \boldmath{$B^{+} \rightarrow p\bar{\Lambda} K^+ K^-$ and $B^{+} \rightarrow \bar{p}\Lambda K^+ K^+$}}

%\input{pub_author_.tex} %Belle official author list
%%% Paper:    B -> p Lambda K K
%%% Journal:  Physical Review (?)
%%% Contacts: ??
%%% Non-responding authors or those who said NO are commented out.
%%% ====================================================================
%%% Click the RELOAD button on your web browser to see the updated file.
%%% ====================================================================
%%% Use \input{author} to insert this material into your latex file.
%%%%% Force institutions to appear in alphabetical order when typeset.
\noaffiliation
\affiliation{University of the Basque Country UPV/EHU, 48080 Bilbao}
\affiliation{Beihang University, Beijing 100191}
%%%\affiliation{University of Bonn, 53115 Bonn}
\affiliation{Brookhaven National Laboratory, Upton, New York 11973}
\affiliation{Budker Institute of Nuclear Physics SB RAS, Novosibirsk 630090}
\affiliation{Faculty of Mathematics and Physics, Charles University, 121 16 Prague}
%%%\affiliation{Chiba University, Chiba 263-8522}
\affiliation{Chonnam National University, Kwangju 660-701}
\affiliation{University of Cincinnati, Cincinnati, Ohio 45221}
\affiliation{Deutsches Elektronen--Synchrotron, 22607 Hamburg}
\affiliation{Duke University, Durham, North Carolina 27708}
%%%\affiliation{University of Florida, Gainesville, Florida 32611}
%%%\affiliation{Department of Physics, Fu Jen Catholic University, Taipei 24205}
\affiliation{Key Laboratory of Nuclear Physics and Ion-beam Application (MOE) and Institute of Modern Physics, Fudan University, Shanghai 200443}
\affiliation{Justus-Liebig-Universit\"at Gie\ss{}en, 35392 Gie\ss{}en}
\affiliation{Gifu University, Gifu 501-1193}
\affiliation{II. Physikalisches Institut, Georg-August-Universit\"at G\"ottingen, 37073 G\"ottingen}
\affiliation{SOKENDAI (The Graduate University for Advanced Studies), Hayama 240-0193}
\affiliation{Gyeongsang National University, Chinju 660-701}
\affiliation{Hanyang University, Seoul 133-791}
\affiliation{University of Hawaii, Honolulu, Hawaii 96822}
\affiliation{High Energy Accelerator Research Organization (KEK), Tsukuba 305-0801}
\affiliation{J-PARC Branch, KEK Theory Center, High Energy Accelerator Research Organization (KEK), Tsukuba 305-0801}
\affiliation{Forschungszentrum J\"{u}lich, 52425 J\"{u}lich}
%%%\affiliation{Hiroshima Institute of Technology, Hiroshima 731-5193}
\affiliation{IKERBASQUE, Basque Foundation for Science, 48013 Bilbao}
%%%\affiliation{University of Illinois at Urbana-Champaign, Urbana, Illinois 61801}
\affiliation{Indian Institute of Science Education and Research Mohali, SAS Nagar, 140306}
\affiliation{Indian Institute of Technology Bhubaneswar, Satya Nagar 751007}
\affiliation{Indian Institute of Technology Guwahati, Assam 781039}
\affiliation{Indian Institute of Technology Hyderabad, Telangana 502285}
\affiliation{Indian Institute of Technology Madras, Chennai 600036}
\affiliation{Indiana University, Bloomington, Indiana 47408}
\affiliation{Institute of High Energy Physics, Chinese Academy of Sciences, Beijing 100049}
\affiliation{Institute of High Energy Physics, Vienna 1050}
\affiliation{Institute for High Energy Physics, Protvino 142281}
%%%\affiliation{Institute of Mathematical Sciences, Chennai 600113}
\affiliation{INFN - Sezione di Napoli, 80126 Napoli}
\affiliation{INFN - Sezione di Torino, 10125 Torino}
\affiliation{Advanced Science Research Center, Japan Atomic Energy Agency, Naka 319-1195}
\affiliation{J. Stefan Institute, 1000 Ljubljana}
\affiliation{Kanagawa University, Yokohama 221-8686}
\affiliation{Institut f\"ur Experimentelle Teilchenphysik, Karlsruher Institut f\"ur Technologie, 76131 Karlsruhe}
%%%\affiliation{Kavli Institute for the Physics and Mathematics of the Universe (WPI), University of Tokyo, Kashiwa 277-8583}
\affiliation{Kennesaw State University, Kennesaw, Georgia 30144}
\affiliation{King Abdulaziz City for Science and Technology, Riyadh 11442}
\affiliation{Department of Physics, Faculty of Science, King Abdulaziz University, Jeddah 21589}
%%%\affiliation{Kitasato University, Tokyo 108-0072}
\affiliation{Korea Institute of Science and Technology Information, Daejeon 305-806}
\affiliation{Korea University, Seoul 136-713}
%%%\affiliation{Kyoto University, Kyoto 606-8502}
\affiliation{Kyungpook National University, Daegu 702-701}
\affiliation{LAL, Univ. Paris-Sud, CNRS/IN2P3, Universit\'{e} Paris-Saclay, Orsay}
\affiliation{\'Ecole Polytechnique F\'ed\'erale de Lausanne (EPFL), Lausanne 1015}
\affiliation{P.N. Lebedev Physical Institute of the Russian Academy of Sciences, Moscow 119991}
\affiliation{Faculty of Mathematics and Physics, University of Ljubljana, 1000 Ljubljana}
\affiliation{Ludwig Maximilians University, 80539 Munich}
\affiliation{Luther College, Decorah, Iowa 52101}
\affiliation{University of Malaya, 50603 Kuala Lumpur}
\affiliation{University of Maribor, 2000 Maribor}
\affiliation{Max-Planck-Institut f\"ur Physik, 80805 M\"unchen}
\affiliation{School of Physics, University of Melbourne, Victoria 3010}
\affiliation{University of Mississippi, University, Mississippi 38677}
\affiliation{University of Miyazaki, Miyazaki 889-2192}
\affiliation{Moscow Physical Engineering Institute, Moscow 115409}
\affiliation{Moscow Institute of Physics and Technology, Moscow Region 141700}
\affiliation{Graduate School of Science, Nagoya University, Nagoya 464-8602}
\affiliation{Kobayashi-Maskawa Institute, Nagoya University, Nagoya 464-8602}
\affiliation{Universit\`{a} di Napoli Federico II, 80055 Napoli}
%%%\affiliation{Nara University of Education, Nara 630-8528}
\affiliation{Nara Women's University, Nara 630-8506}
\affiliation{National Central University, Chung-li 32054}
\affiliation{National United University, Miao Li 36003}
\affiliation{Department of Physics, National Taiwan University, Taipei 10617}
\affiliation{H. Niewodniczanski Institute of Nuclear Physics, Krakow 31-342}
\affiliation{Nippon Dental University, Niigata 951-8580}
\affiliation{Niigata University, Niigata 950-2181}
%%%\affiliation{University of Nova Gorica, 5000 Nova Gorica}
\affiliation{Novosibirsk State University, Novosibirsk 630090}
\affiliation{Osaka City University, Osaka 558-8585}
%%%\affiliation{Osaka University, Osaka 565-0871}
\affiliation{Pacific Northwest National Laboratory, Richland, Washington 99352}
\affiliation{Panjab University, Chandigarh 160014}
\affiliation{Peking University, Beijing 100871}
\affiliation{University of Pittsburgh, Pittsburgh, Pennsylvania 15260}
%%%\affiliation{Punjab Agricultural University, Ludhiana 141004}
%%%\affiliation{Research Center for Electron Photon Science, Tohoku University, Sendai 980-8578}
%%%\affiliation{Research Center for Nuclear Physics, Osaka University, Osaka 567-0047}
\affiliation{Theoretical Research Division, Nishina Center, RIKEN, Saitama 351-0198}
%%%\affiliation{RIKEN BNL Research Center, Upton, New York 11973}
%%%\affiliation{Saga University, Saga 840-8502}
\affiliation{University of Science and Technology of China, Hefei 230026}
%%%\affiliation{Seoul National University, Seoul 151-742}
%%%\affiliation{Shinshu University, Nagano 390-8621}
\affiliation{Showa Pharmaceutical University, Tokyo 194-8543}
\affiliation{Soongsil University, Seoul 156-743}
\affiliation{University of South Carolina, Columbia, South Carolina 29208}
\affiliation{Stefan Meyer Institute for Subatomic Physics, Vienna 1090}
\affiliation{Sungkyunkwan University, Suwon 440-746}
\affiliation{School of Physics, University of Sydney, New South Wales 2006}
\affiliation{Department of Physics, Faculty of Science, University of Tabuk, Tabuk 71451}
\affiliation{Tata Institute of Fundamental Research, Mumbai 400005}
\affiliation{Excellence Cluster Universe, Technische Universit\"at M\"unchen, 85748 Garching}
\affiliation{Department of Physics, Technische Universit\"at M\"unchen, 85748 Garching}
\affiliation{Toho University, Funabashi 274-8510}
%%%\affiliation{Tohoku Gakuin University, Tagajo 985-8537}
\affiliation{Department of Physics, Tohoku University, Sendai 980-8578}
\affiliation{Earthquake Research Institute, University of Tokyo, Tokyo 113-0032}
\affiliation{Department of Physics, University of Tokyo, Tokyo 113-0033}
\affiliation{Tokyo Institute of Technology, Tokyo 152-8550}
\affiliation{Tokyo Metropolitan University, Tokyo 192-0397}
%%%\affiliation{Tokyo University of Agriculture and Technology, Tokyo 184-8588}
%%%\affiliation{Utkal University, Bhubaneswar 751004}
\affiliation{Virginia Polytechnic Institute and State University, Blacksburg, Virginia 24061}
\affiliation{Wayne State University, Detroit, Michigan 48202}
\affiliation{Yamagata University, Yamagata 990-8560}
\affiliation{Yonsei University, Seoul 120-749}
% \author{A.~Abdesselam}\affiliation{Department of Physics, Faculty of Science, University of Tabuk, Tabuk 71451} % Tabuk
  \author{P.-C.~Lu}\affiliation{Department of Physics, National Taiwan University, Taipei 10617} % Taiwan
  \author{M.-Z.~Wang}\affiliation{Department of Physics, National Taiwan University, Taipei 10617} % Taiwan
  \author{R.~Chistov}\affiliation{P.N. Lebedev Physical Institute of the Russian Academy of Sciences, Moscow 119991}\affiliation{Moscow Physical Engineering Institute, Moscow 115409} % Lebedev
  \author{P.~Chang}\affiliation{Department of Physics, National Taiwan University, Taipei 10617} % Taiwan
  \author{I.~Adachi}\affiliation{High Energy Accelerator Research Organization (KEK), Tsukuba 305-0801}\affiliation{SOKENDAI (The Graduate University for Advanced Studies), Hayama 240-0193} % KEK
% \author{K.~Adamczyk}\affiliation{H. Niewodniczanski Institute of Nuclear Physics, Krakow 31-342} % Krakow
  \author{J.~K.~Ahn}\affiliation{Korea University, Seoul 136-713} % Korea
  \author{H.~Aihara}\affiliation{Department of Physics, University of Tokyo, Tokyo 113-0033} % Tokyo
  \author{S.~Al~Said}\affiliation{Department of Physics, Faculty of Science, University of Tabuk, Tabuk 71451}\affiliation{Department of Physics, Faculty of Science, King Abdulaziz University, Jeddah 21589} % Tabuk
% \author{K.~Arinstein}\affiliation{Budker Institute of Nuclear Physics SB RAS, Novosibirsk 630090}\affiliation{Novosibirsk State University, Novosibirsk 630090} % BINP
% \author{Y.~Arita}\affiliation{Graduate School of Science, Nagoya University, Nagoya 464-8602} % Nagoya
  \author{D.~M.~Asner}\affiliation{Brookhaven National Laboratory, Upton, New York 11973} % BNL
  \author{H.~Atmacan}\affiliation{University of South Carolina, Columbia, South Carolina 29208} % SouthCarolina
% ...
% Ljubljana
% \author{N.~Zwahlen}\affiliation{\'Ecole Polytechnique F\'ed\'erale de Lausanne (EPFL), Lausanne 1015} % Lausanne
\collaboration{The Belle Collaboration}

%\date{\today}

\begin{abstract}
We report the study of \OK \ and \SK \ decays using a $772 \times 10^6$ $B\bar{B}$ pair data sample recorded on the $\Upsilon(4S)$ resonance with the Belle detector at KEKB. The following branching fractions are measured: $\mathcal{B}$(\OKPHSP) $=$ $(4.10^{+0.45}_{-0.43}\pm 0.50)\times10^{-6}$, $\mathcal{B}$(\SKPHSP) $=$ $(3.70^{+0.39}_{-0.37} \pm 0.44)\times 10^{-6}$, $\mathcal{B}$(\ETACD+c.c.) $=$ $(2.83^{+0.36}_{-0.34}\pm 0.35)\times 10^{-3}$ and $\mathcal{B}$(\PLPHI) $=$ $(7.95$ $\pm$ $2.09$ $\pm$ $0.77)$ $\times$ $10^{-7}$, where c.c. denotes the corresponding charge-conjugation process. The intermediate resonance decays are excluded in the four-body decay measurements. We also find evidence for $\mathcal{B}$(\ETACS+c.c.) $=$ $(3.48$ $\pm$ $1.48$ $\pm$ $0.46)$ $\times $ $10^{-3}$ and $\mathcal{B}$(\LLKO) $=$ $(2.23$ $\pm$ $0.63$ $\pm$ $0.25)$ $\times$ $10^{-6}$. No significant signals are found for \JPSIS+c.c. and \LLKS; we set the 90\% confidence level upper limits on their decay branching fractions as $< 1.80\times10^{-3}$ and $< 2.08\times10^{-6}$, respectively.
\end{abstract}

%\keywords{Belle; B physics; Baryonic; Threshold effect; Factorization}

\maketitle

Baryonic $B$ decays have been studied at the B-factories~\cite{Bevan2014}, and many intriguing features have been found. Baryon-antibaryon pairs are produced almost
collinearly in most baryonic $B$ decays such that their masses peak
near threshold. There seems to exist a hierarchical structure in the branching fractions of multi-body decays, \textit{e.g.}, $\mathcal{B}$($B^0 \to p\bar{\Lambda}^-_c\pi^+\pi^-$) $>$ $\mathcal{B}$($B^+ \to p\bar{\Lambda}^-_c\pi^+$)
$>$ $\mathcal{B}$($B^0 \to p\bar{\Lambda}^-_c$)~\cite{PDG16}\footnote{Through out this paper, inclusion of charge-conjugate mode is always implied if the charge-conjugate final states are not specifically mentioned together.}. The angular distribution of the proton against the energetic
meson ($K^+$ or $\pi^-$ for the following cases) in the dibaryon system of $B^+ \to p\bar{p}K^+$ and $B^0 \to p\bar{\Lambda}\pi^-$ show a trend opposit to those predicted by theory~\cite{Bevan2014}. These two decays occur presumably via the $b \to s g$ penguin process, where $g$ denotes a hard gluon.

Lately, many more interesting phenomena in baryonic $B$ decays have been found by the LHCb experiment, for example,  very rare two-body decays like $B^0 \to  p \bar{p}$~\cite{PhysRevLett.119.232001}, first evidence for CP violation in baryonic $B$ decays~\cite{Aaij:2014tua}, baryonic $B_s$ decay~\cite{PhysRevLett.119.041802}, baryonic $B_c$ decay~\cite{PhysRevLett.113.152003}, and many first observations of four-body $B^0$ and $B_s$ decays~\cite{PhysRevD.96.051103}.

A generalized factorization picture~\cite{C.Chen08} can qualitatively explain some of the experimental findings. However, the predicted branching fractions may differ by a factor of ten from experimental measurements, \textit{e.g.}, $B^0 \rightarrow p \bar{\Lambda} D^{*-}$ ~\cite{Y.-Y.Chang15}. Later theoretical predictions~\cite{PhysRevD.93.034036} better compare with data after using improved baryonic form factors. It is clear that further studies of baryonic $B$ decays are needed in order to improve theoretical understanding. In this paper, we report measurements of \OK \ and \SK, for which theoretical predictions of $\mathcal{B}$(\OK)~\cite{HSIAO2017348} and $\mathcal{B}$(\PLPHI)~\cite{PhysRevD.85.017501} are available.

The data sample used in this study corresponds to an integrated luminosity of  711 fb$^{-1}$,
which contains $772\times10^{6}$ $B\bar{B}$ pairs produced at the $\Upsilon(4S)$ resonance.
The Belle detector~\cite{Abashian02,Brodzicka12} is located at the interaction point (IP) of
the  KEKB asymmetric-energy $e^{+}$ (3.5 GeV) $e^{-}$ (8 GeV) collider~\cite{Kurokawa03,Abe13}.
It is a large-solid-angle spectrometer comprising six specialized subdetectors:
the Silicon Vertex Detector, the 50-layer Central Drift Chamber (CDC),
the Aerogel Cherenkov Counter (ACC), the Time-Of-Flight scintillation counter (TOF),
the electromagnetic calorimeter (ECL), and the $K_L^0$ and muon detector (KLM).
A superconducting solenoid surrounding all but the KLM produces a $1.5$ T magnetic field.

In this analysis, we combine $p\bar{\Lambda} K^+ K^-$ ($\bar{p}\Lambda K^+ K^+$) to form $B^+$ candidates. We require charged particles (tracks from $\Lambda$ are excluded) to originate near the IP, less than $1.0$ cm away along the positron beam direction and less than $0.2$ cm away in the transverse plane. To identify a kaon or a proton track,
we use the likelihood information from the charged-hadron identification system (CDC, ACC, TOF)~\cite{NAKANO2002402} and apply the same selection criteria as in Ref.~\cite{ppk}. We use information from ECL and KLM to reject charged particles resembling electrons and muons. We require $\Lambda$($p\pi^-$) candidates to have a displaced vertex that is consistent with a long-lived particle originating from the IP and a mass between 1.111 and 1.121 GeV/$c^2$.
%To further suppress background, we require that the positive \Lambda daughter should be consistent with
%a proton.

We use the following two variables, $\Delta E \equiv E_{\rm recon} - E_{\rm beam}$ and $M_{\rm bc} \equiv \sqrt{(E_{\rm beam}/c^2)^2-(P_{\rm recon}/c)^2}$, to identify signal, where $E_{\rm recon}$/$P_{\rm recon}$ and $E_{\rm beam}$ are the reconstructed $B$ energy/momentum and beam energy measured in the $\Upsilon(4S)$ rest frame, respectively. We define 5.24 $< M_{\rm bc} <$ 5.29 GeV$/c^2$ and $|\Delta E| <$ 0.2 GeV as the fit region; 5.27 $< M_{\rm bc} <$ 5.29 GeV$/c^2$ and $|\Delta E| <$ 0.03 GeV as the signal region.

%We use the following two kinematic variables with information measured at the $\Upsilon(4S)$ center-of-mass frame
%to identify the reconstructed $B^+$ candidates: the energy
%difference $\Delta E=E_{\rm recon} - E_{\rm beam}$ and the beam-energy-constrained mass
%$M_{\rm bc}c^2 = \sqrt{(E_{\rm beam})^2-(P_{\rm recon})^2c^2}$. Here $E_{\rm beam}$, $E_{\rm recon}$, and $P_{recon}$
%are the beam energy, the reconstructed $B^+$ energy, and the reconstructed $B^+$ momentum, respectively.
%We define a sample box satisfying $M_{\rm bc} >$ 5.24 GeV$/c^2$ and $|\Delta E| <$ 0.2 GeV to extract
%signal yield by the maximum likelihood fit method; we also define a signal box satisfying 5.273
%$< M_{\rm bc} <$ 5.287 GeV$/c^2$ and $|\Delta E| <$ 0.03 GeV to make projection plots.

The dominant background is from the continuum process ($e^+e^- \rightarrow q\bar{q}$, $q=u,d,s,c$).
We generate phase space \OKPHSP \ and \SKPHSP \ signal events and continuum background
using EvtGen~\cite{EvtGen} and later process them with a GEANT3-based detector simulation program
that provides the detector-level information~\cite{Geant}.
These Monte Carlo (MC) samples are used to optimize the signal selection criteria. We use a neural network
package, Neurobayes~\cite{NB}, for background suppression. There are 21 input variables for the training of Neurobayes: 17 modified Fox-Wolfram moments treating the information of particles involved in the signal $B$ candidate separately from those in the rest of the event~\cite{KSFW, KSFW2} to distinguish spherical $B \bar{B}$ events from the jet-like $q\bar{q}$ events, the missing mass of each event, the vertex difference between the $B^+$ candidate and the accompanying $B$, the angle between $B^+$ flight direction and the beam axis in the $\Upsilon(4S)$ rest frame, and the tagging information for the accompanying $B$ ~\cite{qr}. The output value of Neurobayes is between $+1$ ($B\bar{B}$-like) and $-1$ ($q\bar{q}$-like). The optimized selection and its related systematic uncertainty is mode dependent.

We consider at most one $B^+$ candidate in each event: if there are multiple candidates, we select the one with the smallest ($\chi^2_{B\_{\rm vtx}}+\chi^2_{\Lambda\_{\rm vtx}}$), where $\chi^2_{B(\Lambda)\rm\_vtx}$  represents the $\chi^2$ value of $B$($\Lambda$) vertex fit. The probability to have multiple $B$ candidates is less than 6\% and the success rate of this selection is larger than 92\% according to MC study.

In the investigation of possible intermediate states in \OK \ and \SK, we check the mass spectra from combinations of various final-state particles in and near the signal region. We find many intermediate resonances: $\eta_c$, $J/\psi$ and $\chi_{c1}$ in $M(p\bar{\Lambda}K^-)$; $\phi$ in $M(K^+K^-)$; $\Lambda(1520)$ in $M(p K^-)$.
After removing events in the mass windows of resonances: 2.92 $<M(p\bar{\Lambda}K^-)<$ 3.11 GeV/$c^2$ for $\eta_c$ and $J/\psi$, 3.49 $<M(p\bar{\Lambda}K^-)<$ 3.53 GeV/$c^2$ for $\chi_{c1}$, 1.01 $<M(K^+K^-)<$ 1.03 GeV/$c^2$ for $\phi$, and 1.46 $<M(pK^-)<$ 1.58 GeV/$c^2$ for $\Lambda(1520)$,
we still observe a large number of signal events. We attribute them to genuine four-body decays. Note that there is no significant $D^0$ peak found. We also find a threshold peak mixed with the phase space distribution in the $p\bar{\Lambda}$ mass spectrum. Therefore, we generate signal MC samples with this feature to mimic data. This mixing ratio is mode dependent in order to match with data.

We use an extended unbinned maximum likelihood fit to extract signal yields of genuine \OK \ and \SK \ four-body decays. The likelihood function is defined as

$$\mathcal{L}=\frac{e^{-(N_{s}+N_{b})}}{N!}\prod_{i=1}^{N} (N_{s} P_{s}( \Delta{E}^i, M_{\rm bc}^i) +N_{b} P_{b}(  \Delta{E}^i, M_{\rm bc}^i)),$$
where $N$ is the number of total events, $i$ denotes the event index, $N_{s}$ and $N_{b}$ are fit parameters representing the numbers of signal events and background events, respectively; $P_s$ and $P_b$ are the probability density functions of signal and background, respectively.

\begin{figure}[htbp]
\footnotesize
\centering
\includegraphics[width=4.2cm]{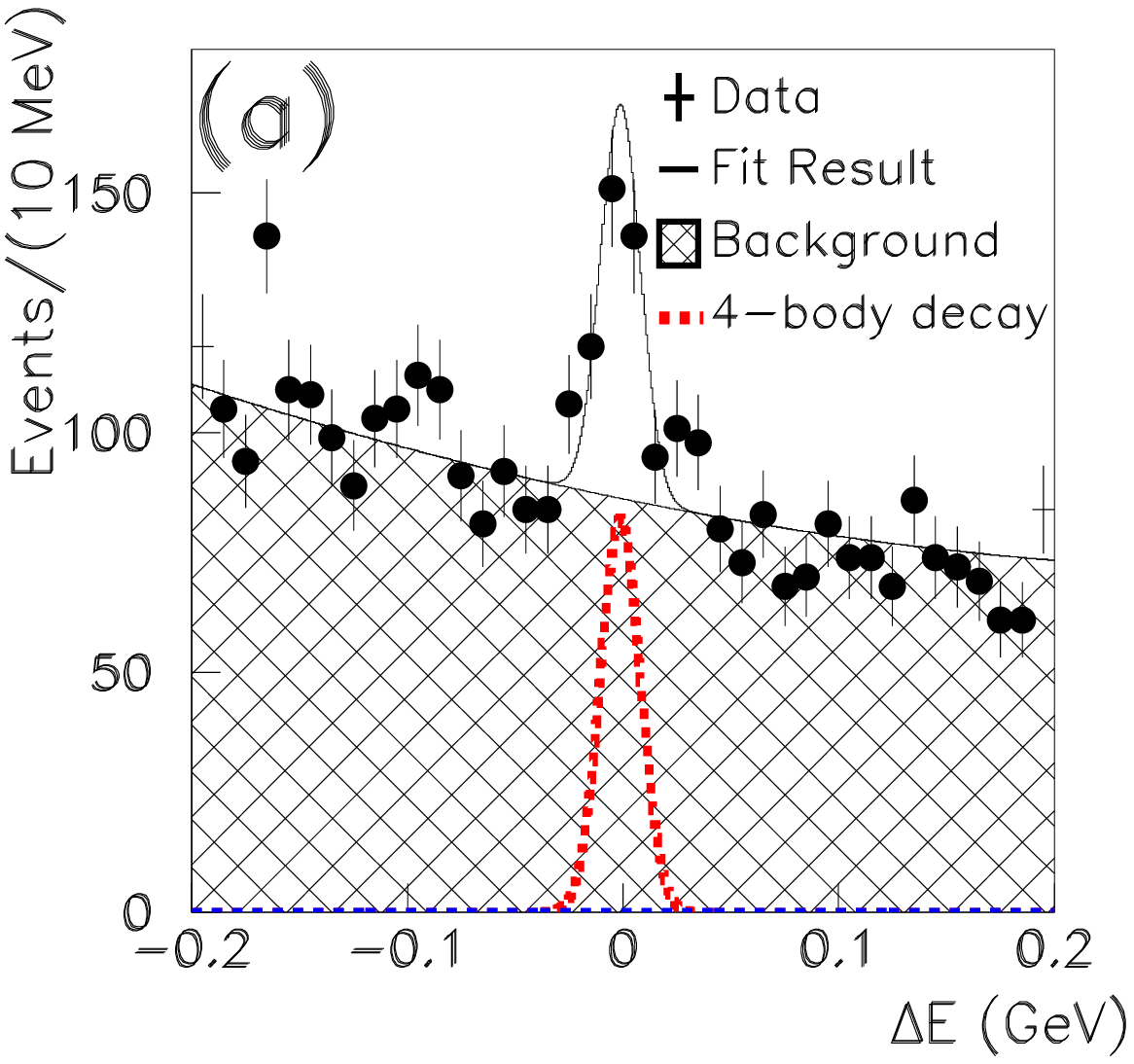}
\includegraphics[width=4.2cm]{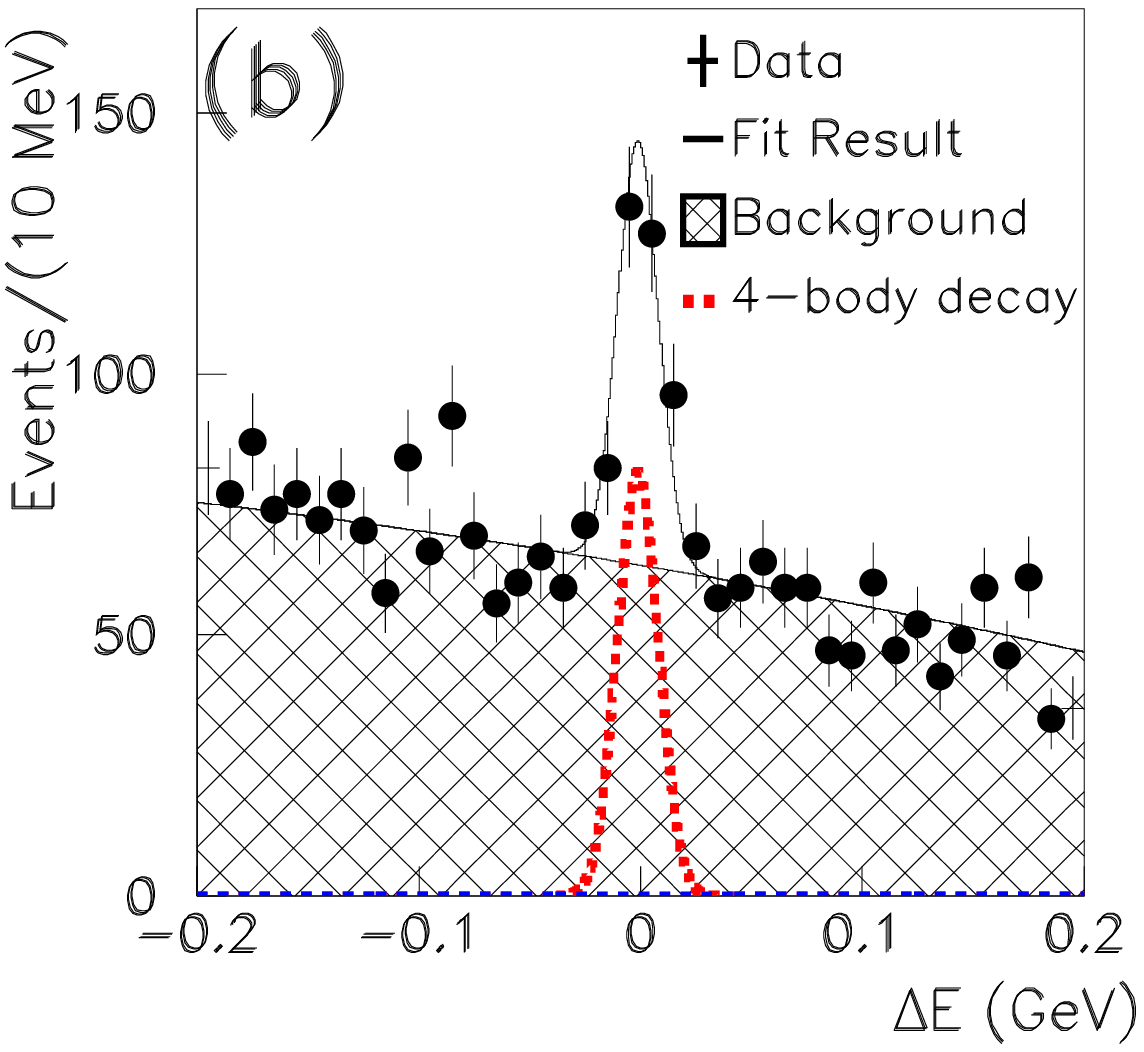}
\includegraphics[width=4.2cm]{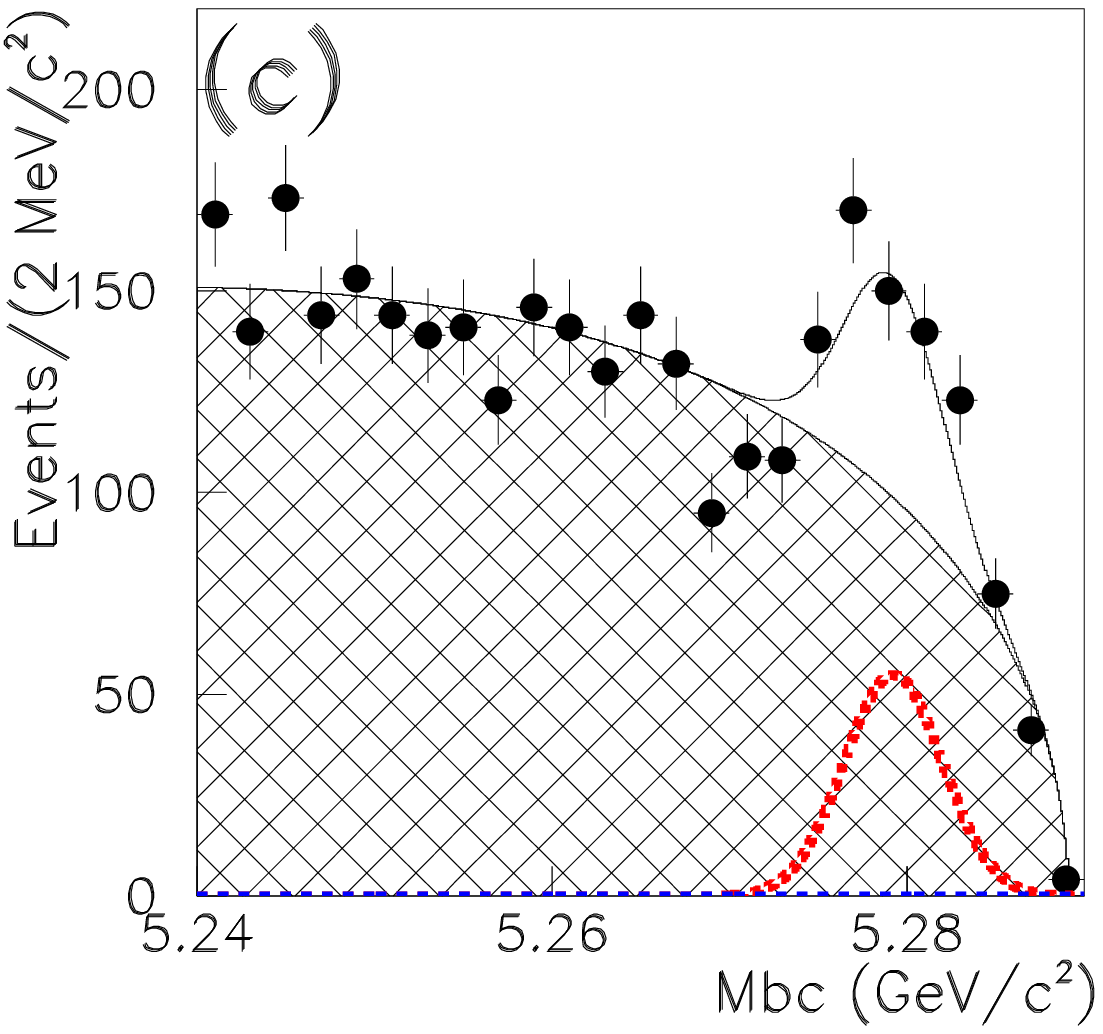}
\includegraphics[width=4.2cm]{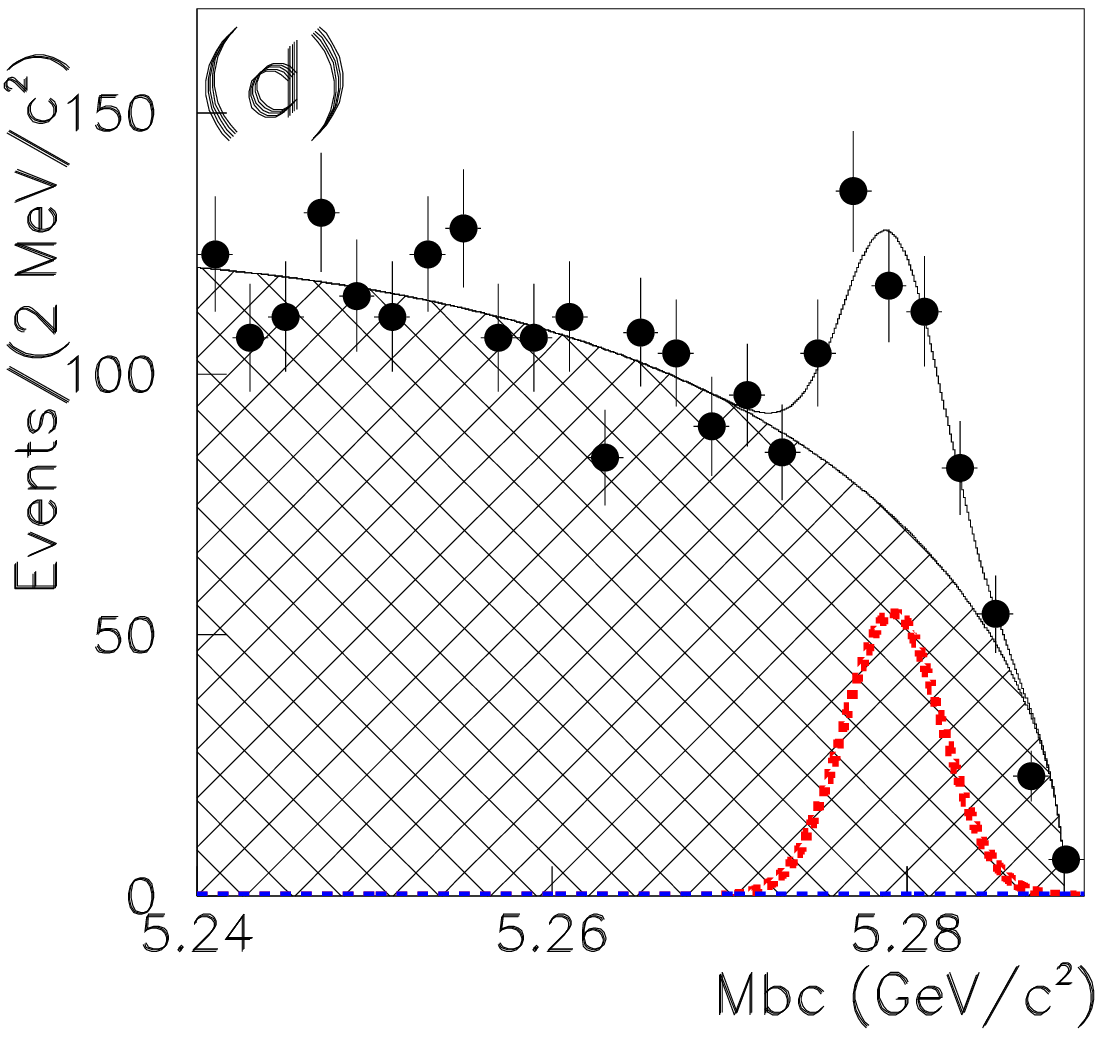}
\caption{Fit results of genuine four-body decays in projection plots of $\Delta E$ (5.27 $<M_{bc}<$ 5.29 \GeV) and $M_{bc}$ ($|\Delta E|<$ 0.03 GeV). (a)(c) are for the final state $p \bar{\Lambda} K^+ K^-$; (b)(d) are for the final state $\bar{p} \Lambda K^+ K^+$.}
\label{fig:2dfitting}
\end{figure}

\begin{figure}[htbp]
\footnotesize
\centering
{
\includegraphics[width=6cm]{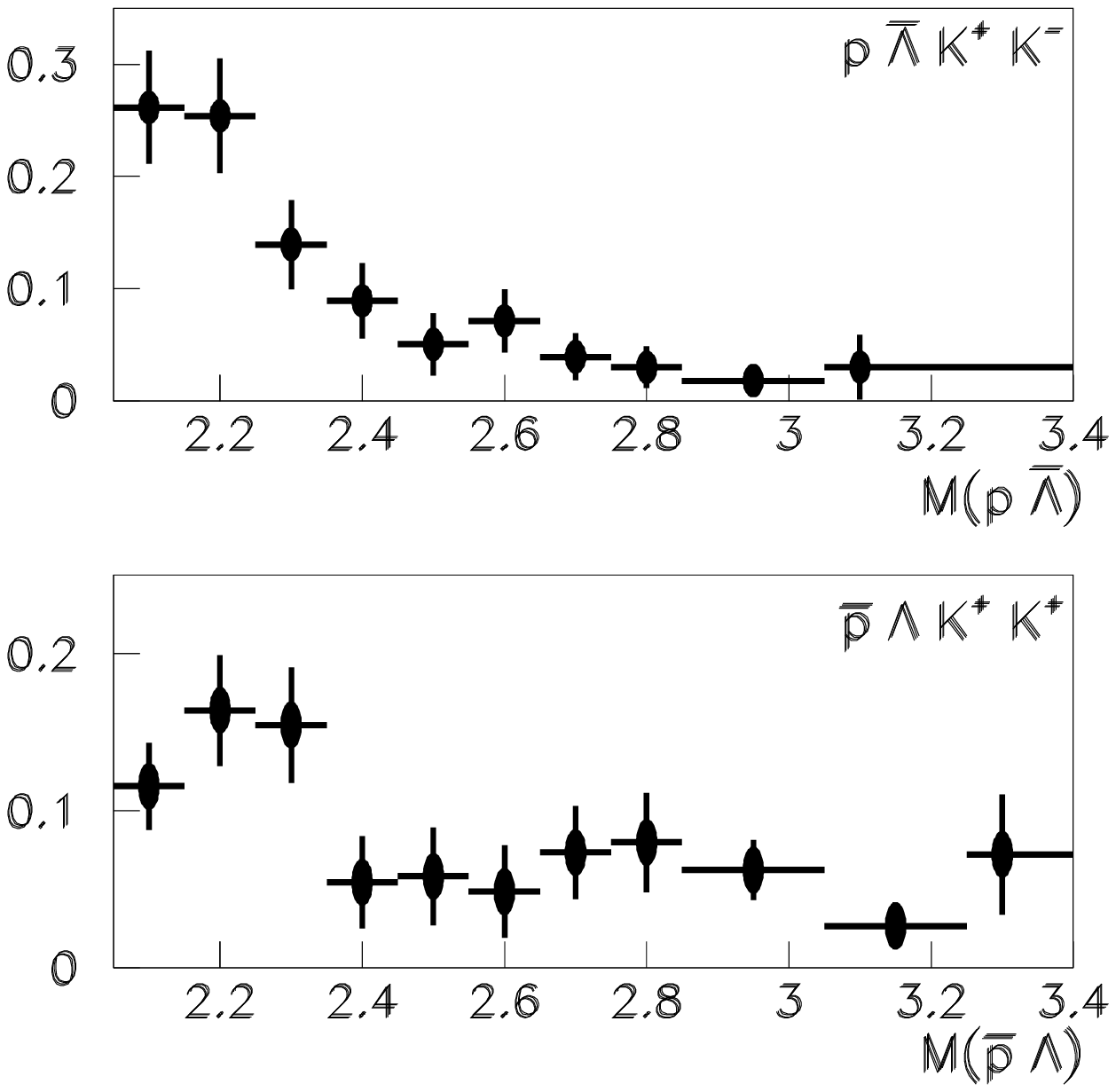}
}
\caption{Normalized and efficiency-corrected signal yield distributions of M($p \bar{\Lambda}$) and M($\bar{p}\Lambda$) for four-body decays. Clear threshold peaks are observed.}
\label{fig:fittingByBin}
\end{figure}

Backgrounds like generic ($b \to c$) $B$ decays and other rare ($b \to u,d,s$) $B$ decays, after investigation of MC simulation, show no peak in the fit region. We combine them with continuum background as the general background to fit with. We use Gaussian functions to model the signal shapes in both $\Delta E$ and $M_{\rm bc}$, a second-order polynomial function for the background $\Delta E$ distribution and an ARGUS function~\cite{argus} for the background $M_{\rm bc}$ distribution. The fit results are displayed in Fig.~\ref{fig:2dfitting}. Note that the possible feed-down events from $B^+ \to p\bar{\Sigma^0}K^+K^-$ and $B^+ \to \bar{p}\Sigma^0 K^+K^+$ will form a peak around $-0.1$ GeV in the $\Delta{E}$ spectra. The fit bias due to this excess around $-0.1$ GeV is negligible ($<0.4\%$). We apply the same fitting procedure in bins of $M_{p\bar{\Lambda}/\bar{p}\Lambda}$ to determine the signal yields. The corresponding normalized and efficiency-corrected signal yield distributions are shown in Fig.~\ref{fig:fittingByBin}. Clear threshold peaks and non-negligible phase space contributions are observed.

%Points with error bars are data, the dotted line is signal, the hatched region is background, the solid black curve is the total distribution of all components.

\begin{figure}[htbp]
\centering
{
\includegraphics[width=4.2cm]{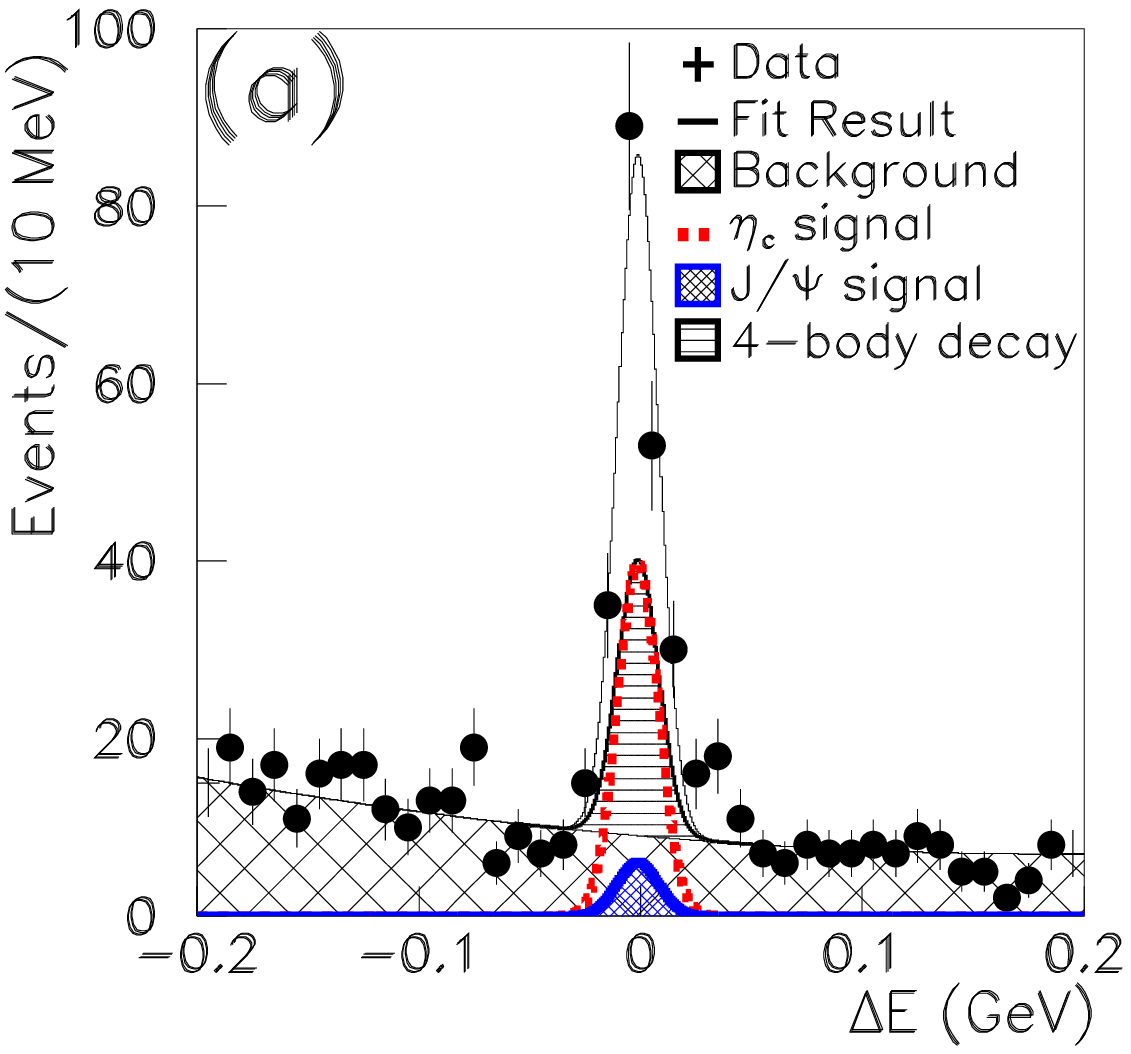}
\includegraphics[width=4.2cm]{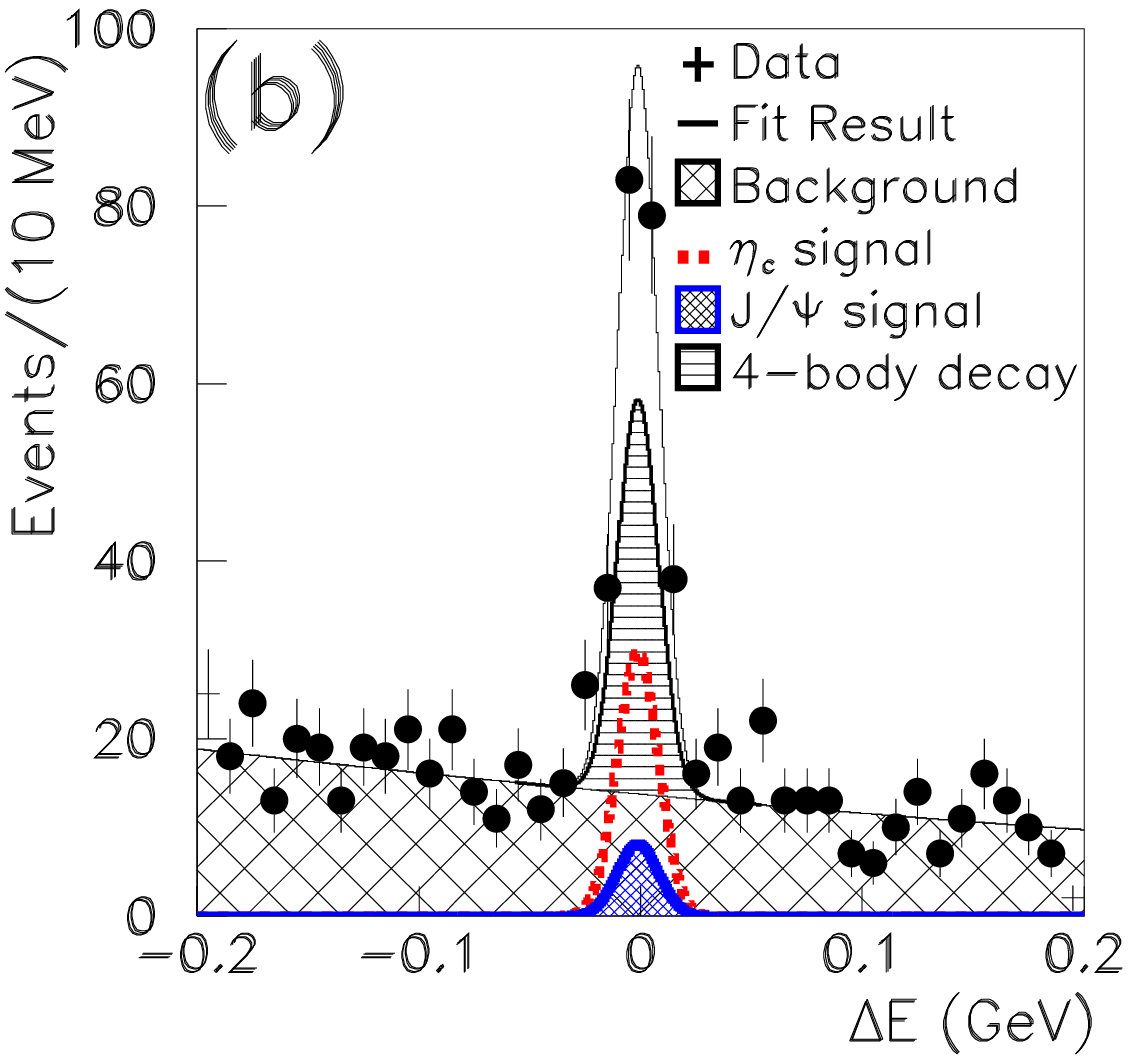}
\includegraphics[width=4.2cm]{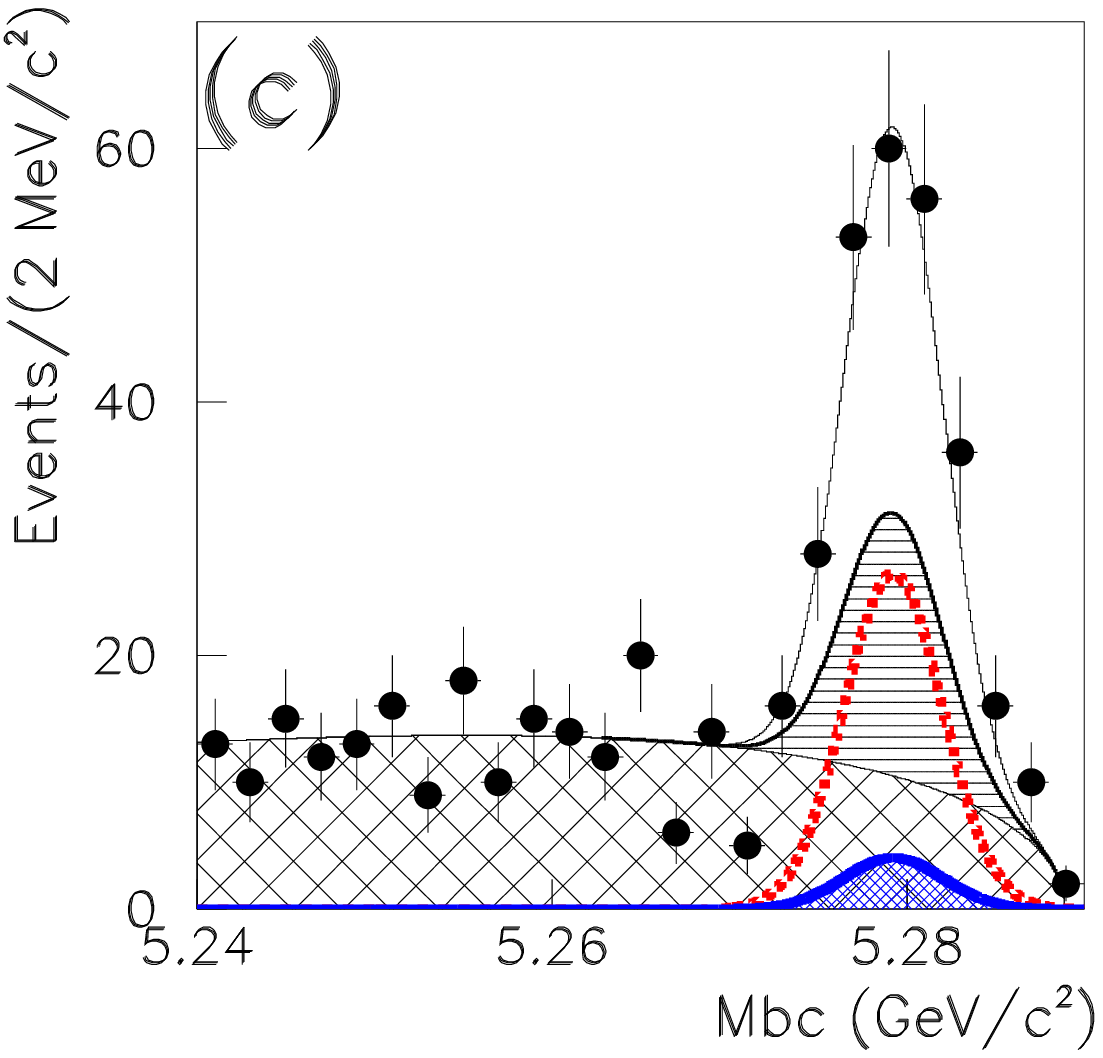}
\includegraphics[width=4.2cm]{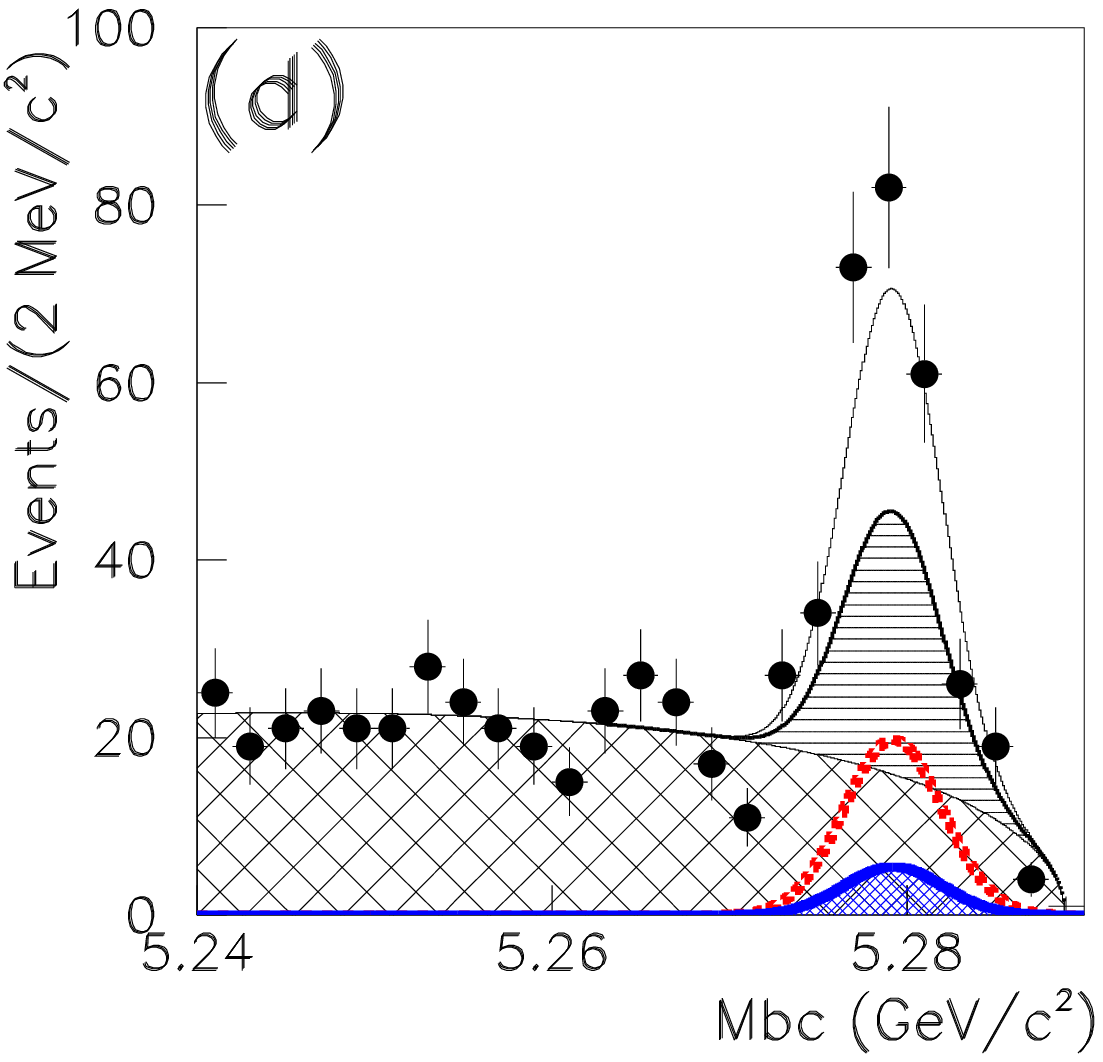}
\includegraphics[width=4.2cm]{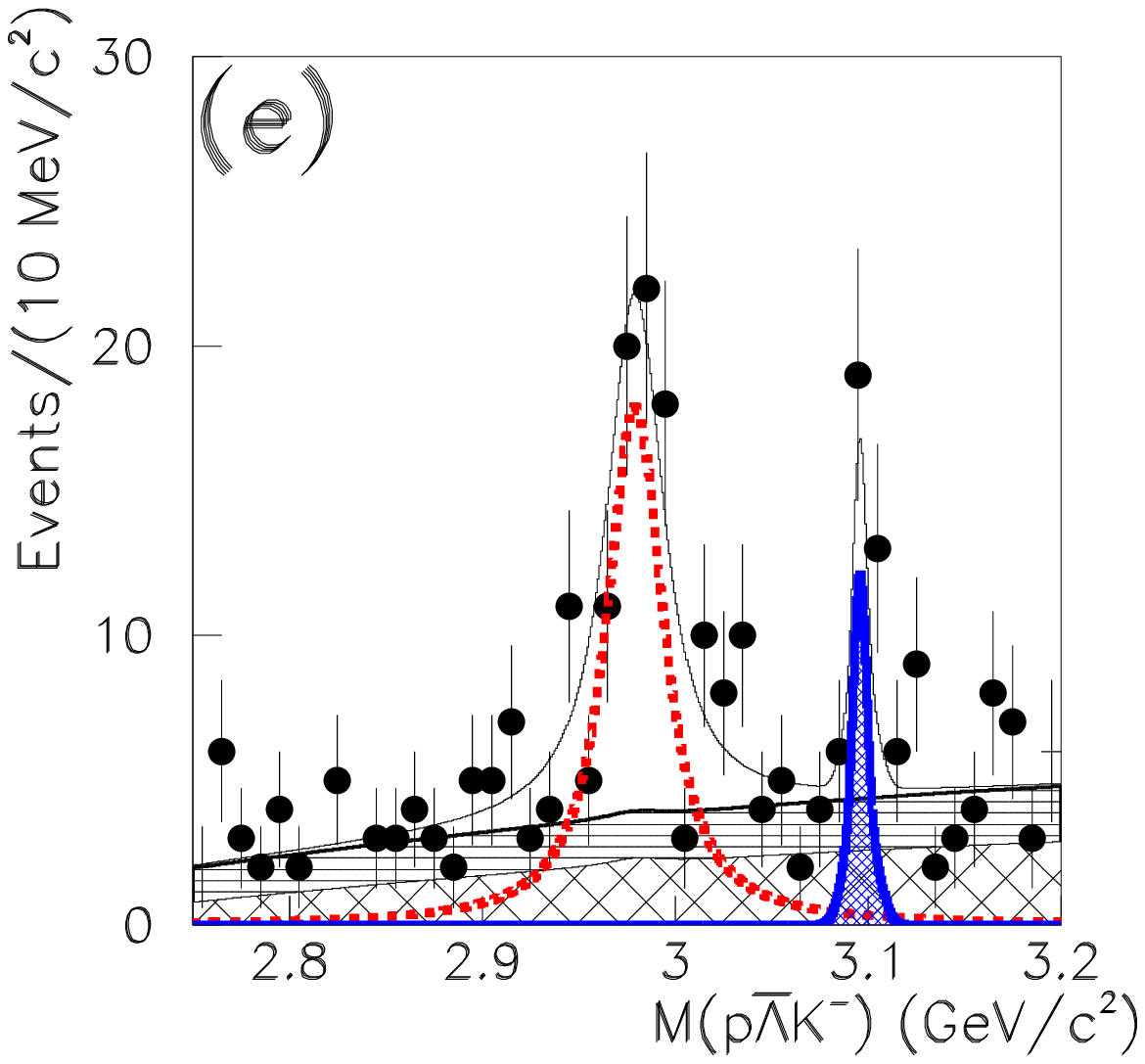}
\includegraphics[width=4.2cm]{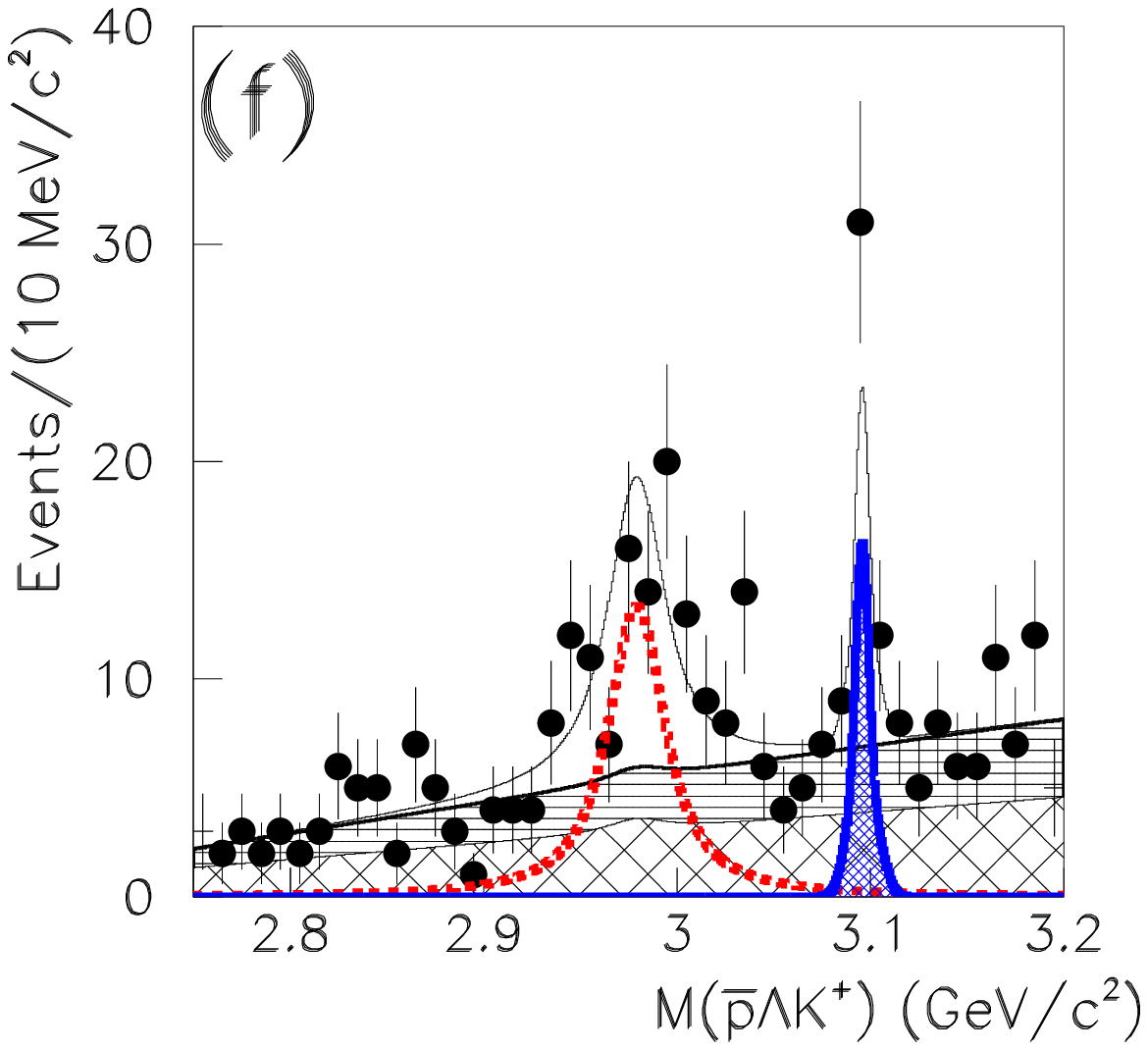}
}
\caption{Fit results of \ETAC(\ETACD) and \JPSI(\JPSID) with 2.75 $<M_{p\bar{\Lambda}K^-}/M_{\bar{p}\Lambda K^+}<$ 3.2 \GeV \ in projection plots of $\Delta E$ (5.27 $<M_{bc}<$ 5.29 \GeV), $M_{bc}$ ($|\Delta E|<$ 0.03 GeV) and $M_{p\bar{\Lambda}K^-}/M_{\bar{p} \Lambda K^+}$ (in signal box). (a)(c)(e) are for the final state $p \bar{\Lambda} K^+ K^-$; (b)(d)(f) are for the final state $\bar{p} \Lambda K^+ K^+$. For illustration purpose, we only show signal curve peaking in all spectra and four-body decay as horizontal-line region, and  merge all backgrounds as cross-hatched region.}
\label{fig:fitting3dcc}
\end{figure}

\begin{figure}[htbp]
\centering
{
\includegraphics[width=4.2cm]{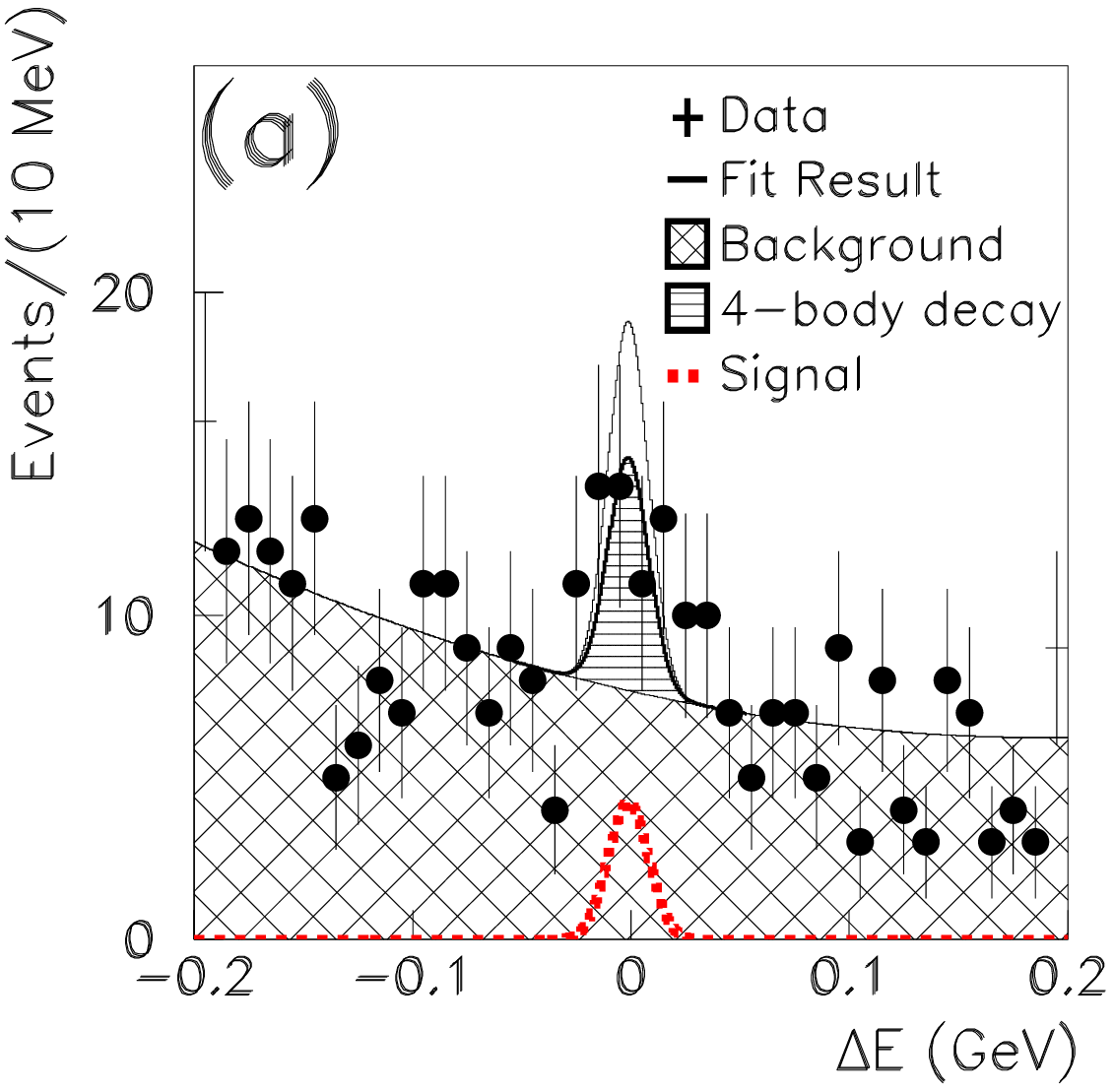}
\includegraphics[width=4.2cm]{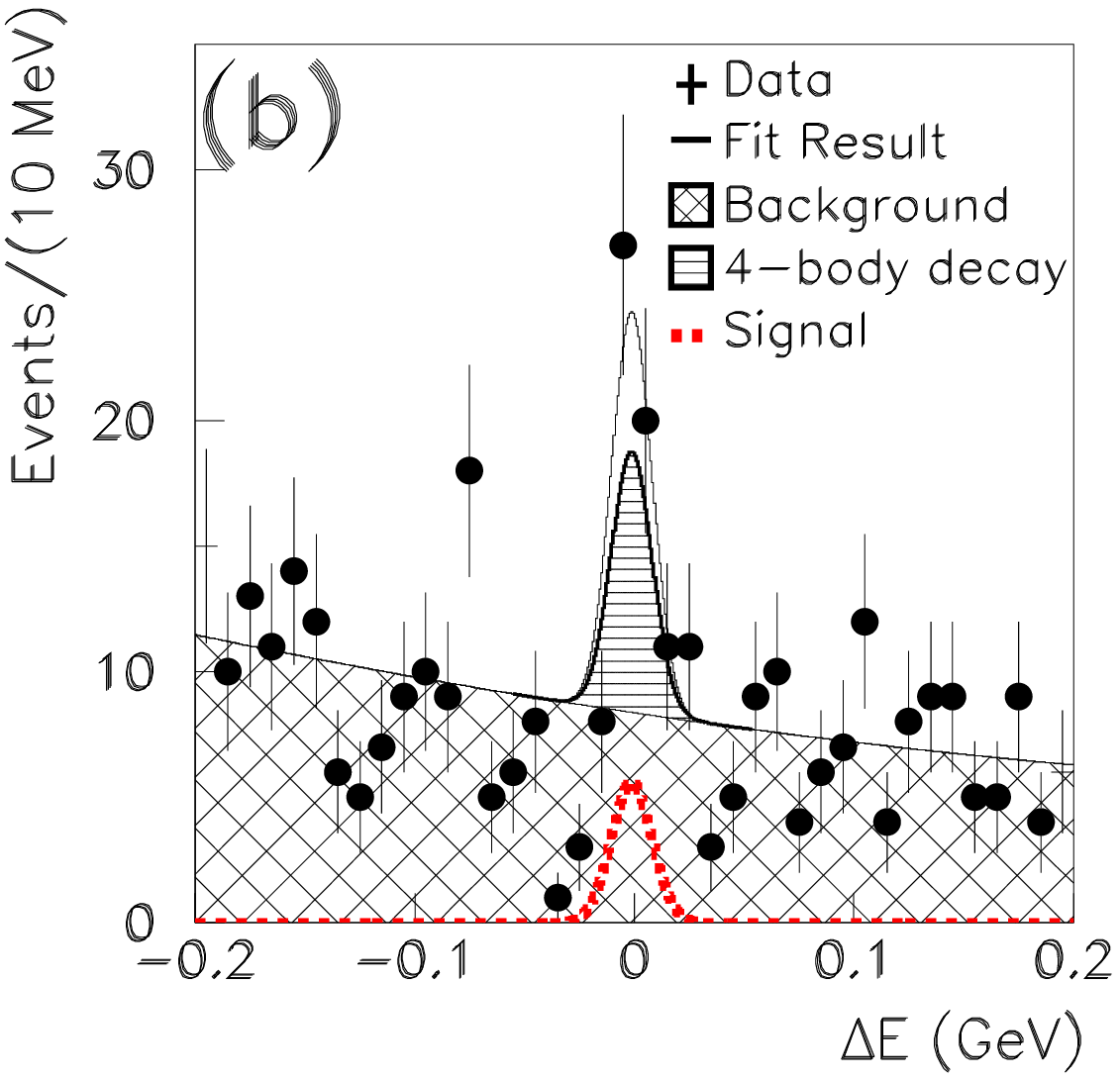}
\includegraphics[width=4.2cm]{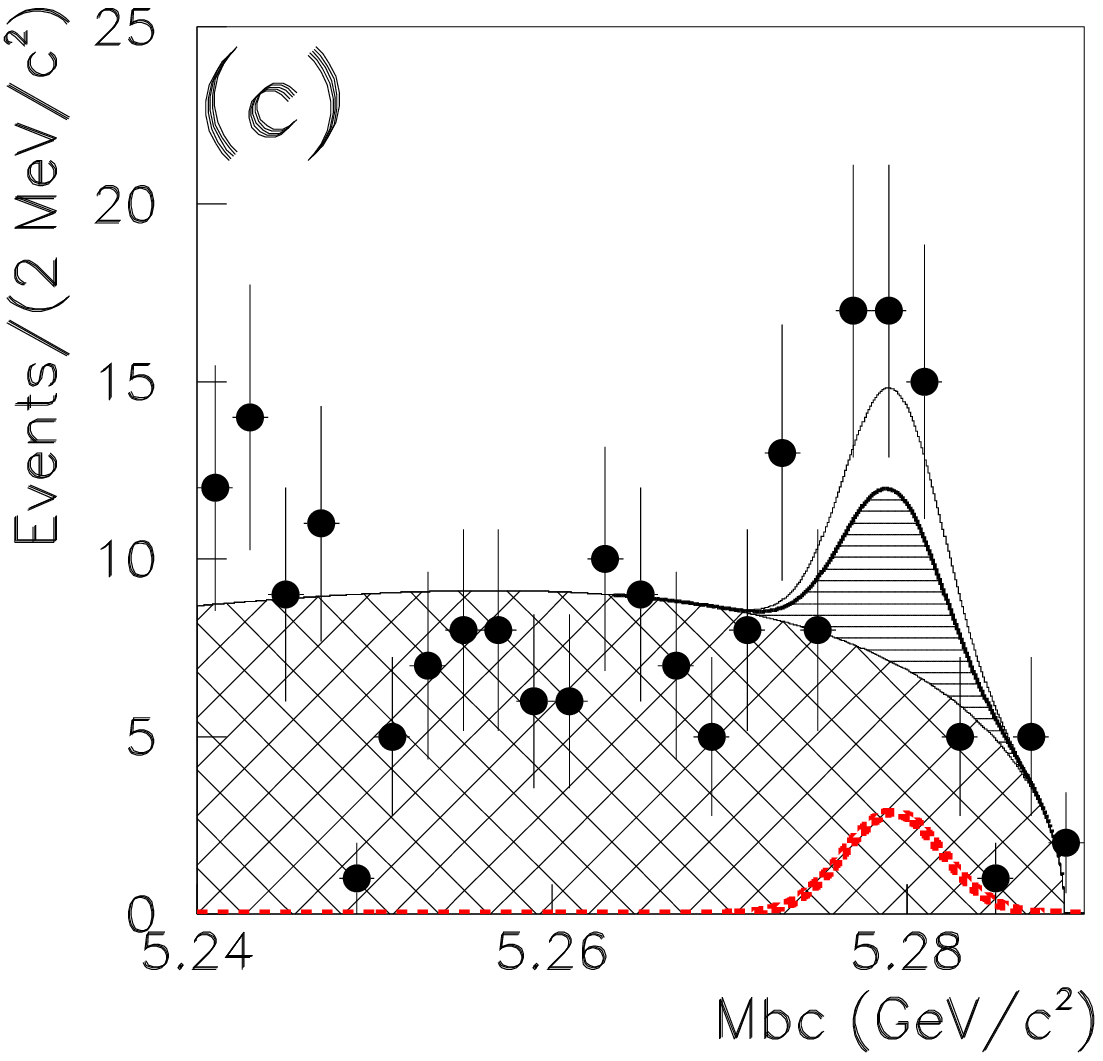}
\includegraphics[width=4.2cm]{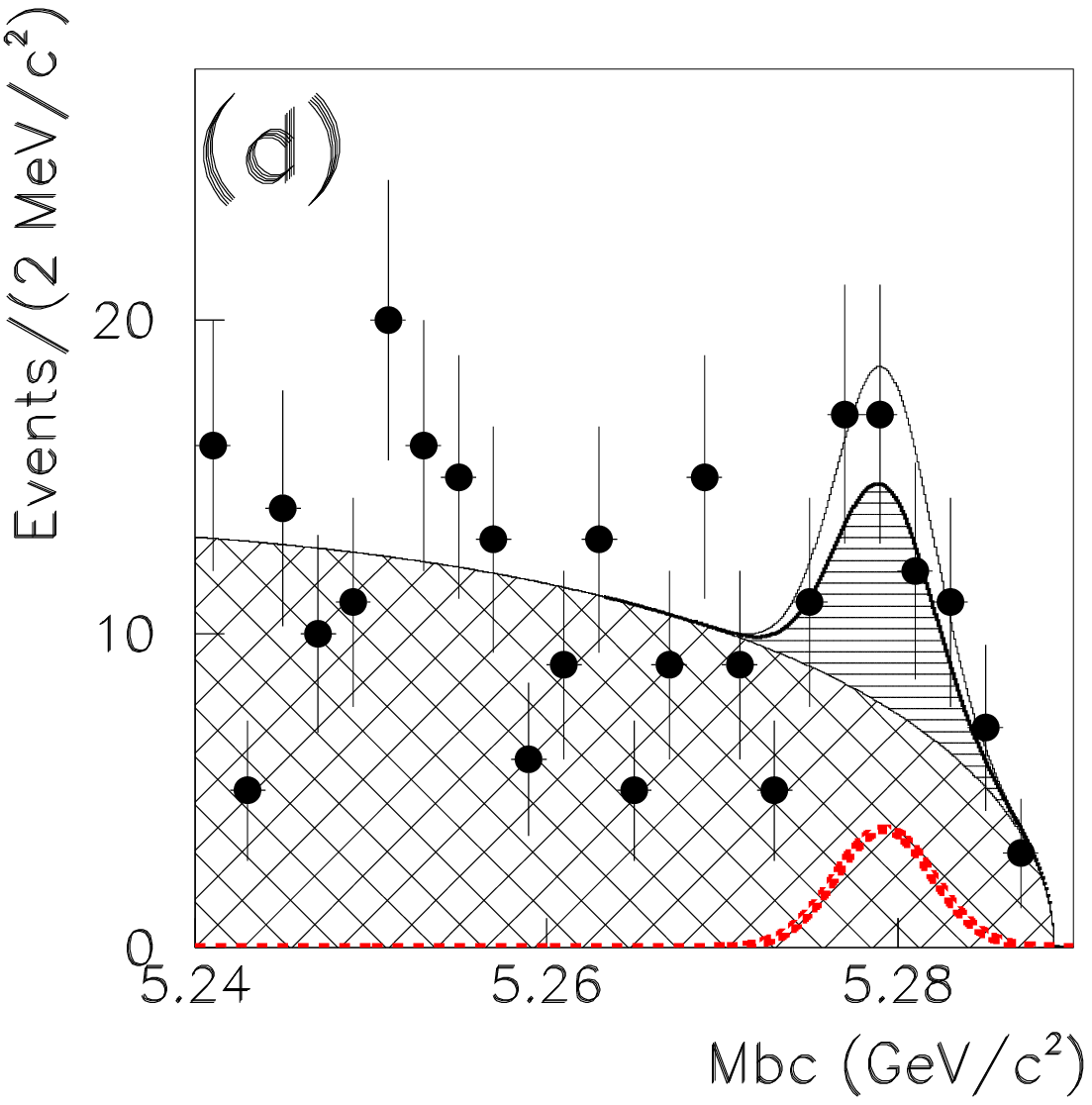}
\includegraphics[width=4.2cm]{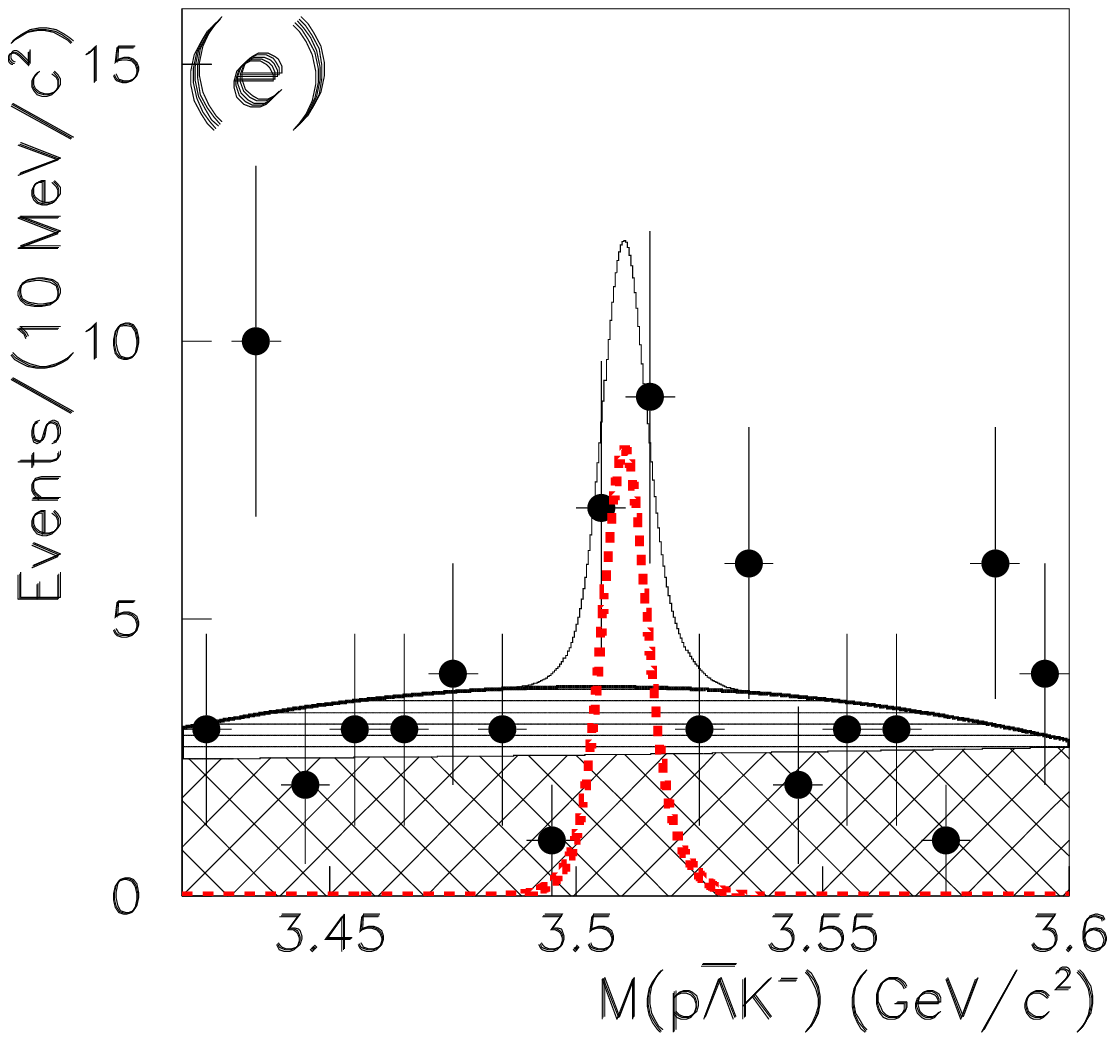}
\includegraphics[width=4.2cm]{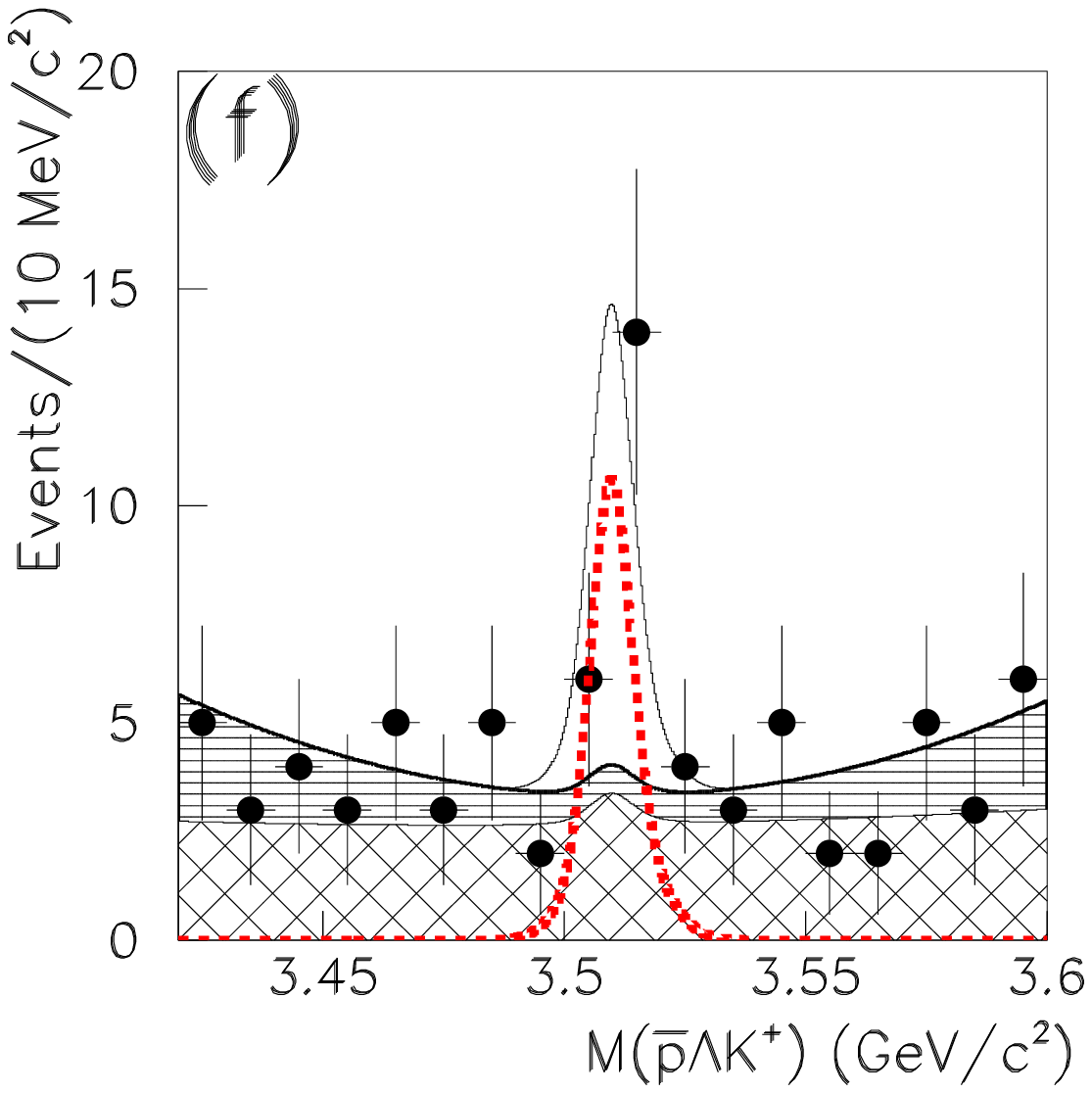}
}
\caption{Fit results of \CHIC \ with 3.42 $<M_{p\bar{\Lambda} K^-}/M_{\bar{p}\Lambda K^+}<$ 3.6 \GeV \ in projection plots of  $\Delta E$ (5.27 $<M_{bc}<$ 5.29 \GeV), $M_{bc}$ ($|\Delta E|<$ 0.03 GeV) and $M_{p\bar{\Lambda} K^-}/M_{\bar{p}\Lambda K^+}$ (in signal box). (a)(c)(e) are for the final state $p \bar{\Lambda} K^+ K^-$; (b)(d)(f) are for the final state $\bar{p} \Lambda K^+ K^+$. For illustration purpose, we only show signal curve
peaking in all spectra and four-body decay as horizontal-line region, and  merge all backgrounds as cross-hatched region.}
\label{fig:fitting3dchic1}
\end{figure}

\begin{figure}[htbp]
\centering
{
\includegraphics[width=4.2cm]{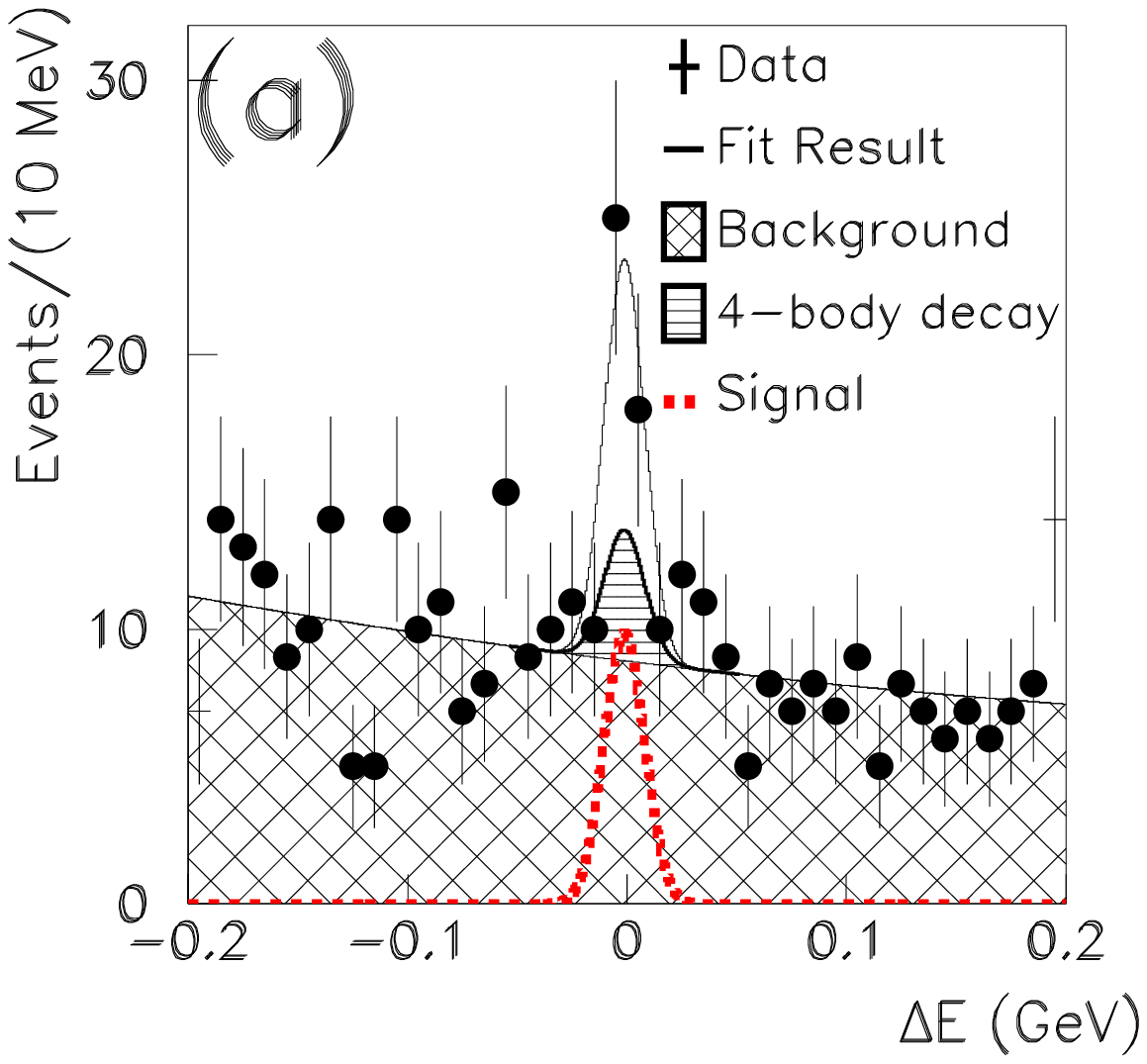}
\includegraphics[width=4.2cm]{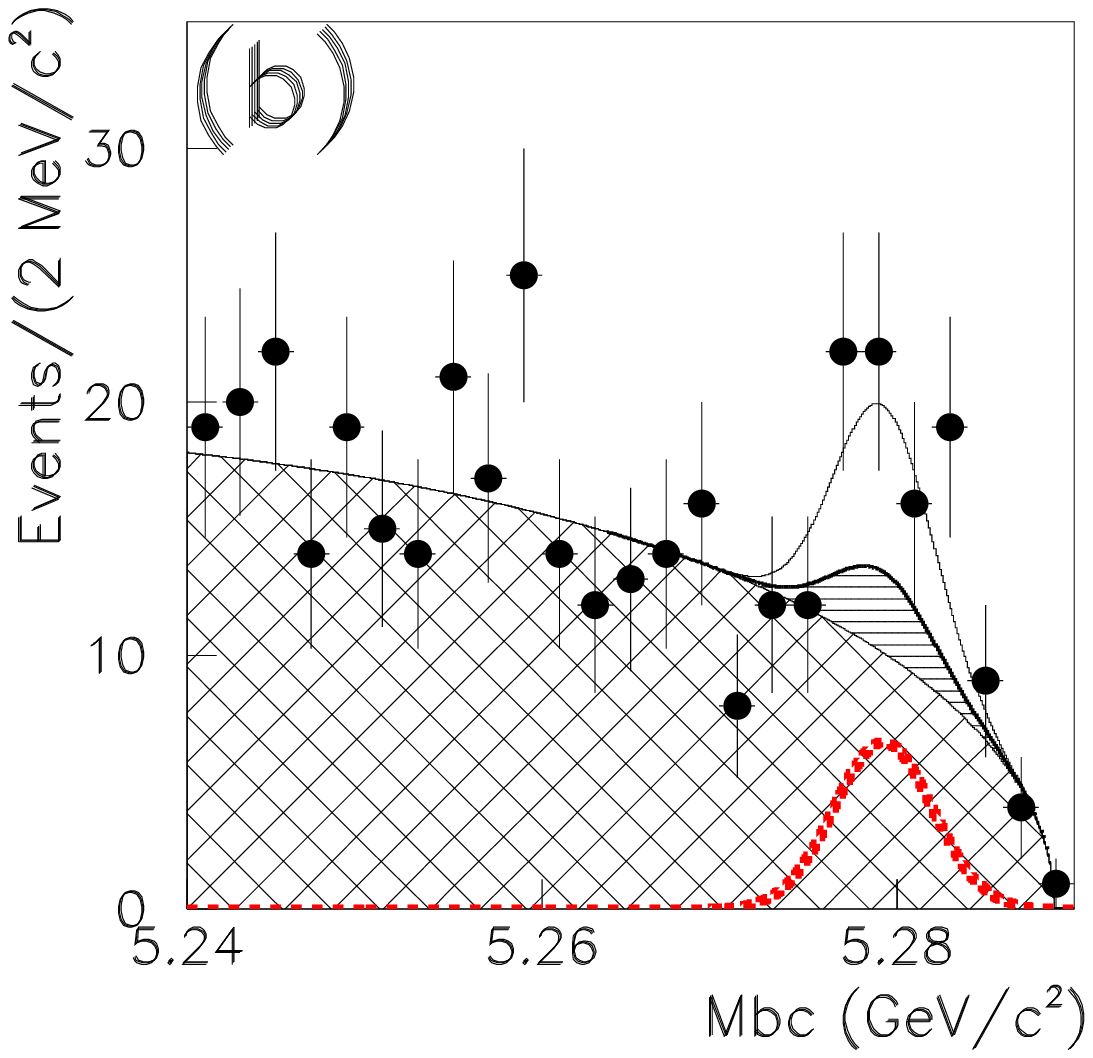}
\includegraphics[width=4.2cm]{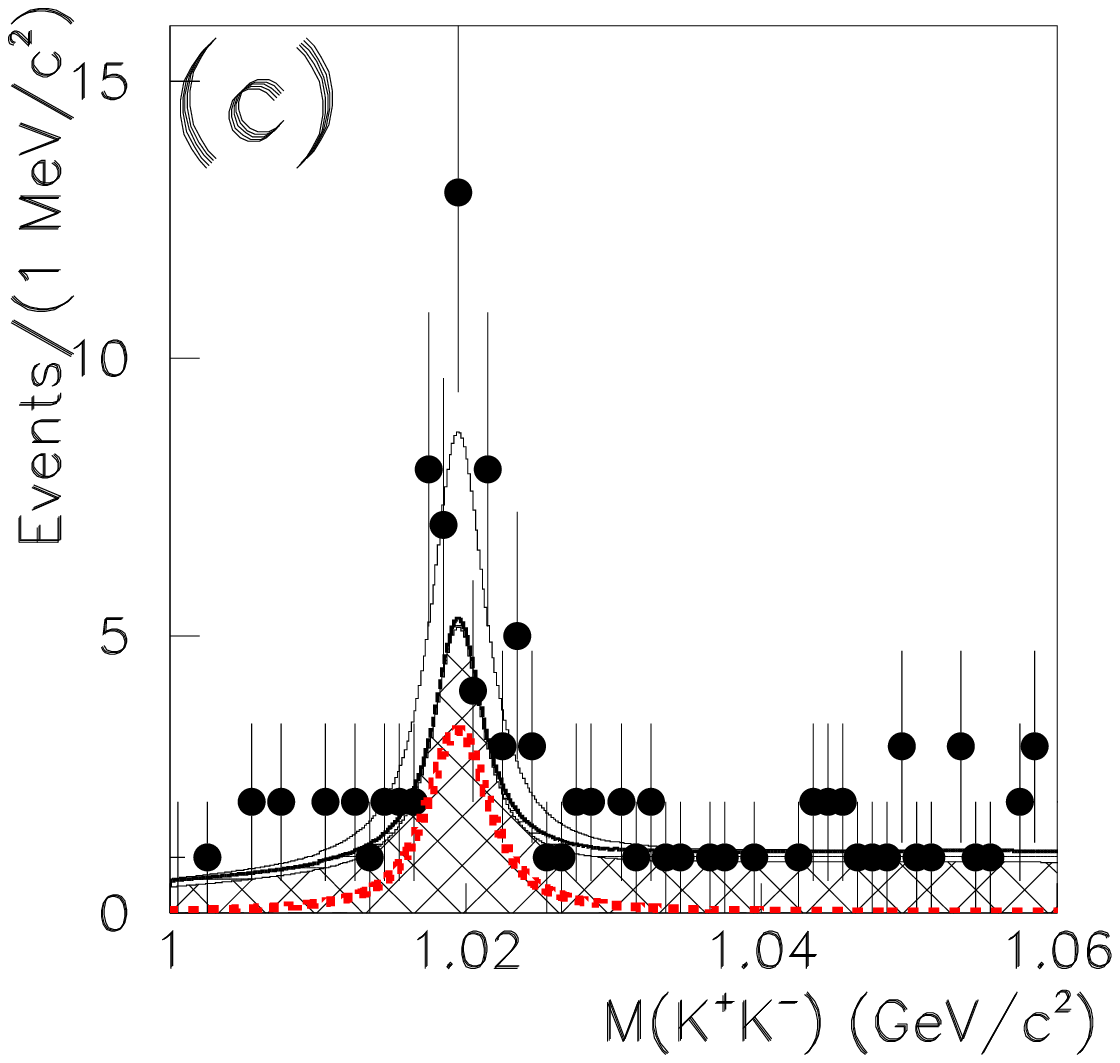}
}
\caption{Fit result of $B^+ \rightarrow  p\bar{\Lambda}\phi$ with 1.00 $<M_{\rm K^+K^-}<$ 1.08 \GeV \ in projection plots of $\Delta E$ (5.27 $<M_{bc}<$ 5.29 \GeV), $M_{bc}$ ($|\Delta E|<$ 0.03 GeV) and $M_{K^+K^-}$ (in signal box). For illustration purpose, we only show signal curve
peaking in all spectra and four-body decay as horizontal-line region, and  merge all backgrounds as cross-hatched region.}
\label{fig:fitting3d4}
\end{figure}

\begin{figure}[htbp]
\centering
{
\includegraphics[width=4.2cm]{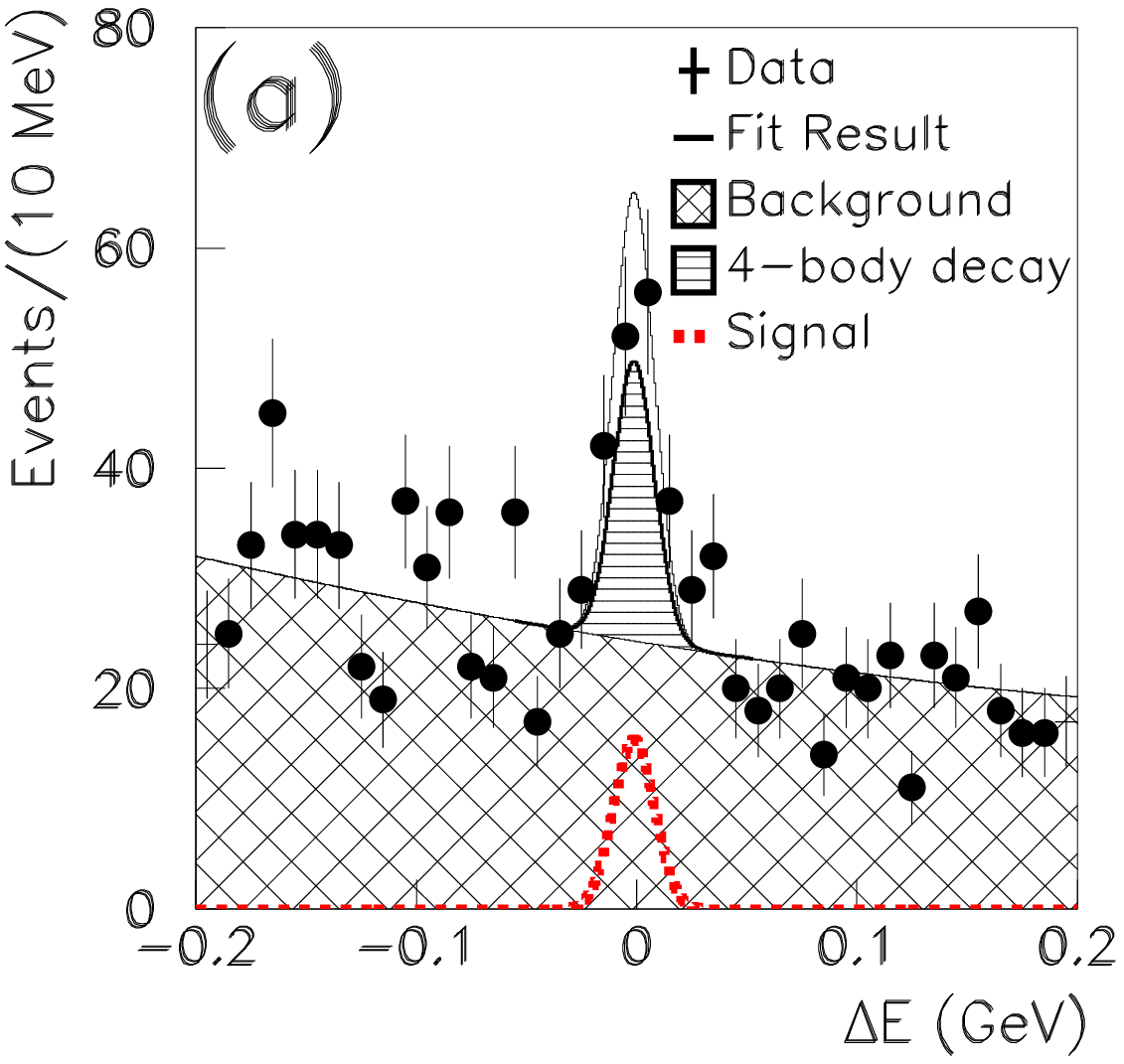}
\includegraphics[width=4.2cm]{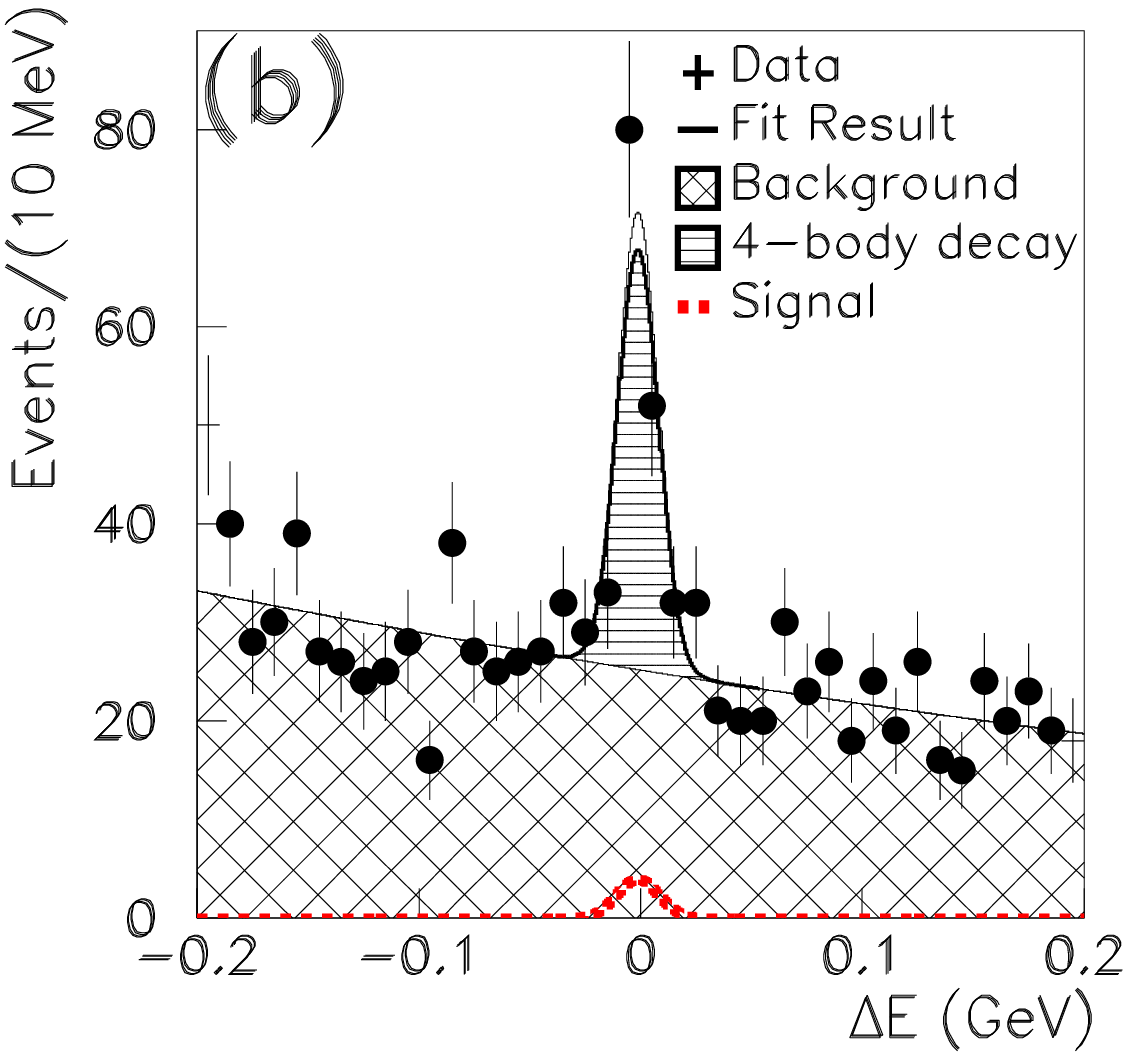}
\includegraphics[width=4.2cm]{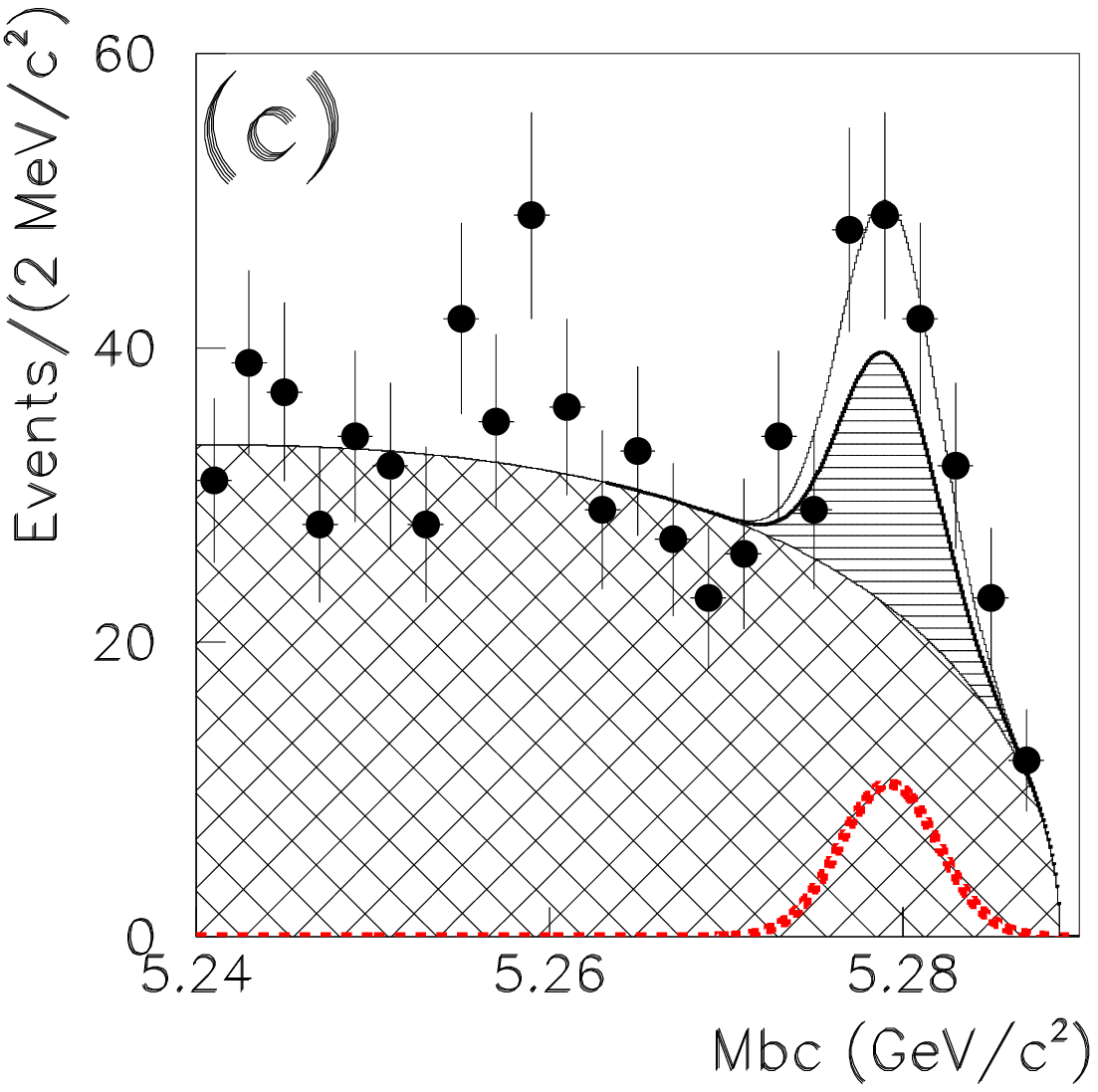}
\includegraphics[width=4.2cm]{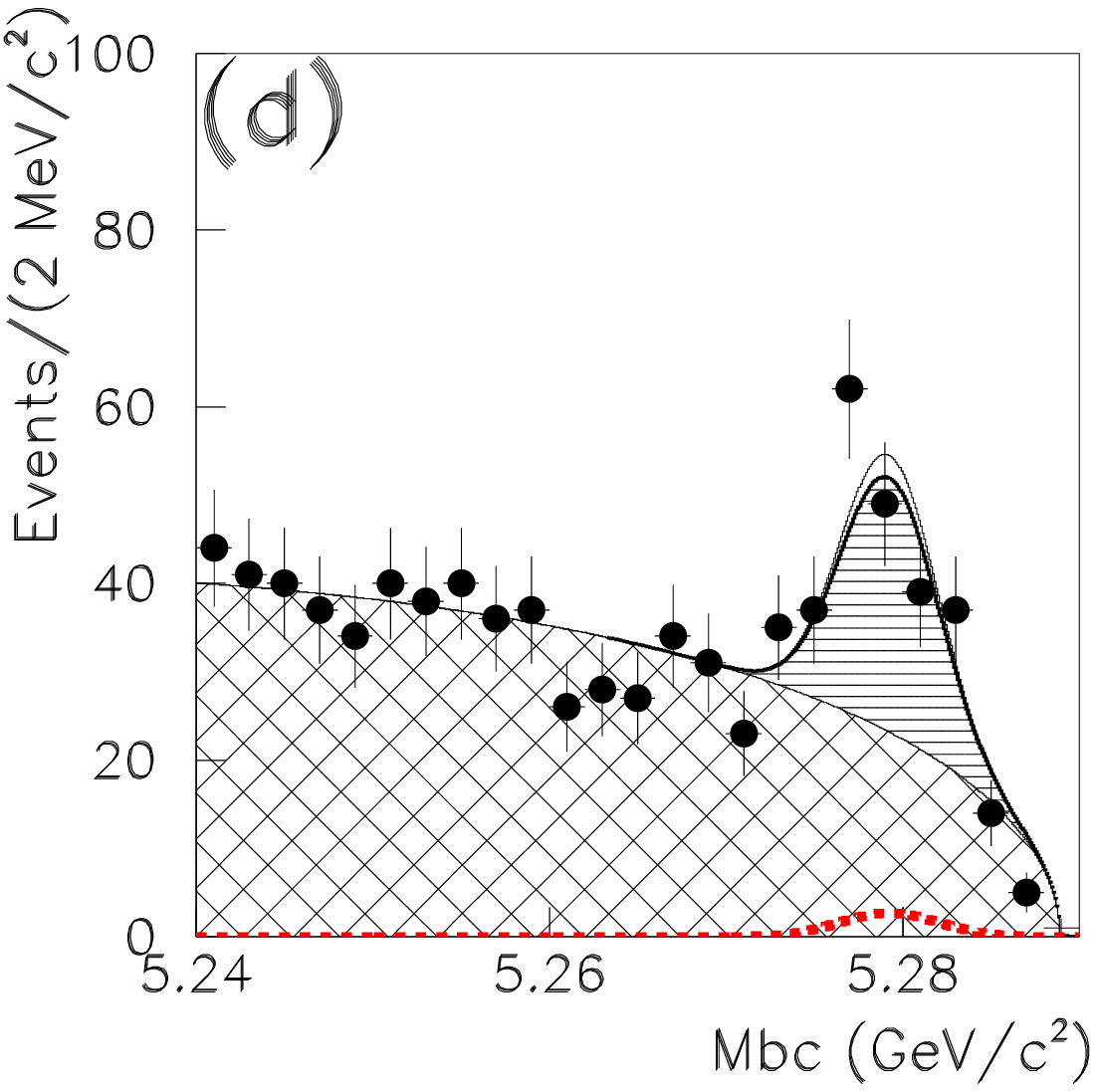}
\includegraphics[width=4.2cm]{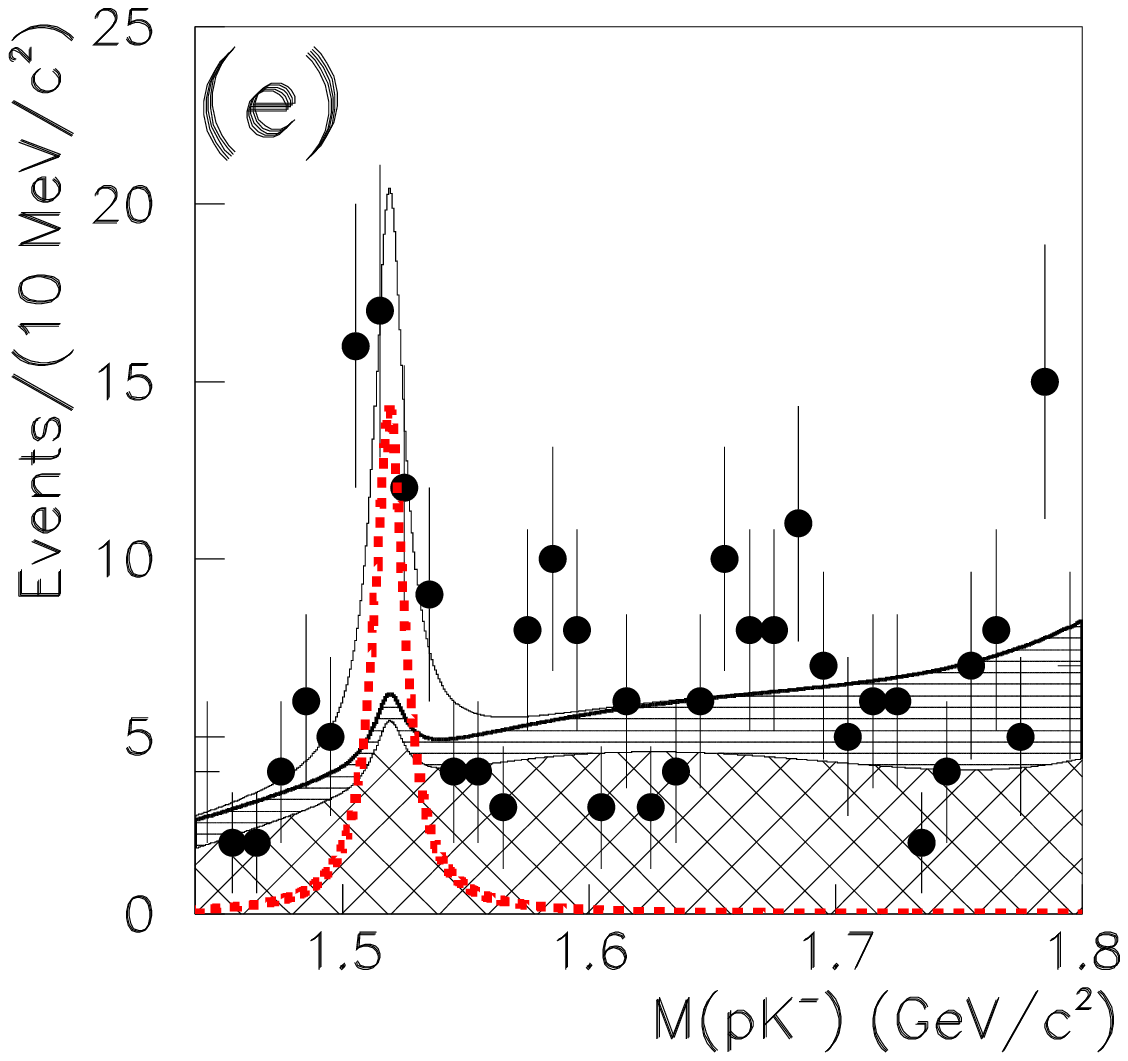}
\includegraphics[width=4.2cm]{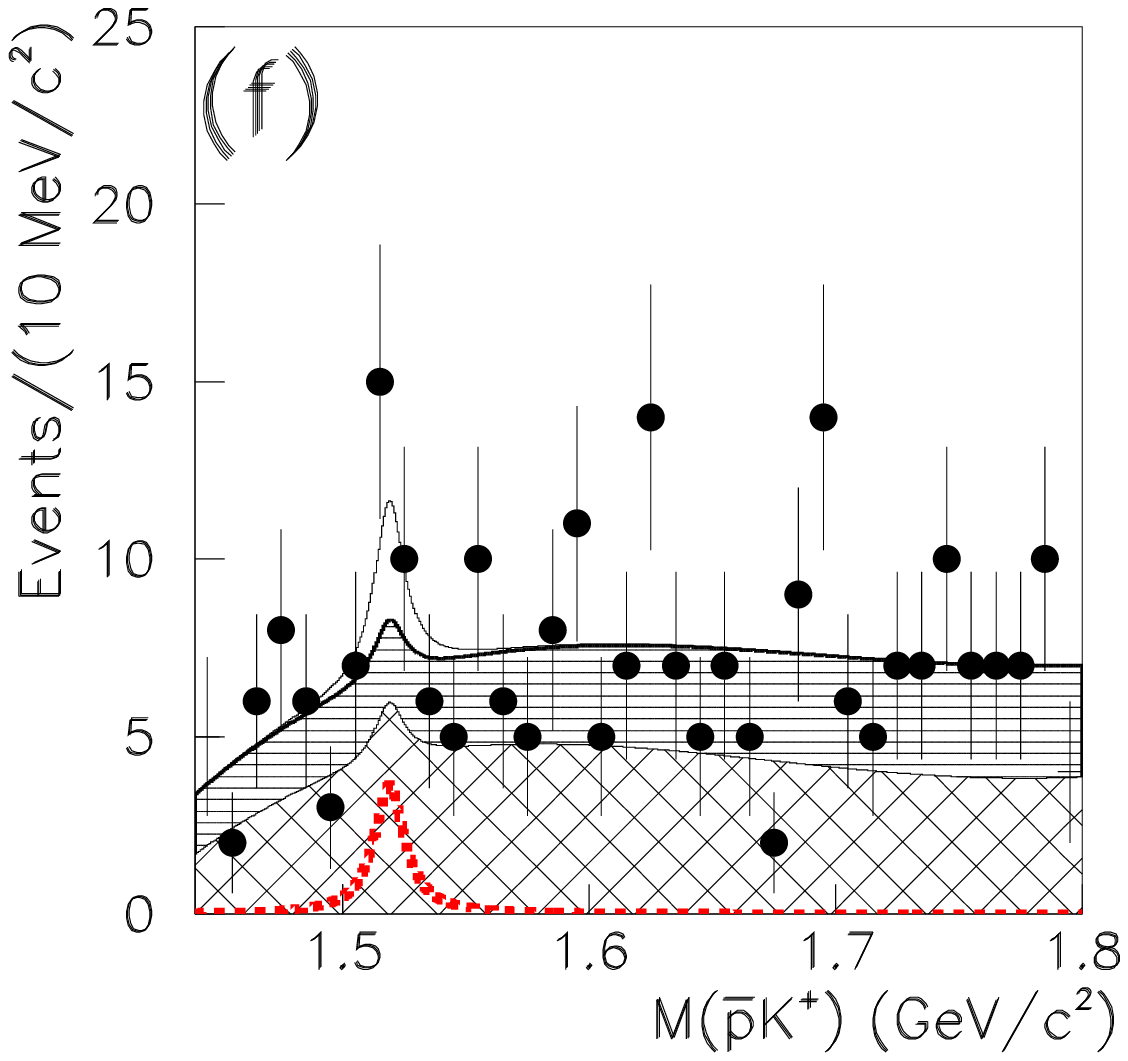}
}
\caption{Fit results of \LLKO \ and \LLKS \ with 1.44 $<M_{pK^-}/M_{\bar{p}K^+}<$ 1.8 \GeV \ in projection plots of  $\Delta E$ (5.27 $<M_{bc}<$ 5.29 \GeV), $M_{bc}$ ($|\Delta E|<$ 0.03 GeV) and $M_{p K^-}/M_{\bar{p} K^+}$ (in signal box). (a)(c)(e) are for the final state $p \bar{\Lambda} K^+ K^-$; (b)(d)(f) are for the final state $\bar{p} \Lambda K^+ K^+$. For illustration purpose, we only show signal curve
peaking in all spectra and four-body decay as horizontal-line region, and  merge all backgrounds as cross-hatched region.}
\label{fig:fitting3dL}
\end{figure}

Since the signal yield is significant enough, we fix the signal shapes in a similar likelihood fit to extract the signal yields with intermediate resonances $\eta_c$, $J/\psi$, $\chi_{c1}$, $\Lambda(1520)$ and $\phi$. In addition to $\Delta E$ and  $M_{\rm bc}$, we include the invariant mass of an intermediate resonance as a third variable in our fit assuming that the probability density function, $P(M_{\rm res})$, is independent of $P(\Delta E, M_{\rm bc})$. We use the world average mass and width values of these resonances to generate MC samples~\cite{PDG16}. For $\eta_c$ and $\phi$, we use a Breit-Wigner function convolved with a Gaussian function; for $J/\psi$ and $\chi_{c1}$, we use the sum of two Gaussian functions in order to fit the corresponding MC mass distributions; for $\Lambda(1520)$, we use one Breit-Wigner function. The obtained signal shapes are fixed in the later data fit. We use a 2nd-order polynomial function to model the background shape in the resonance mass spectrum. The different components of the fit function are the resonance signal (peaking in all spectra), genuine four-body signal (only peaking in  $\Delta E$ and $M_{\rm bc}$), background with resonances produced by other processes (only peaking in  $M_{\rm res}$) and non-peaking background. In contrast to fixed peaking shapes, all non-peaking shapes are floated and determined from the fit. Figure~\ref{fig:fitting3dcc} shows the fit results for \ETAC \ (\ETACD) and \JPSI \ (\JPSID). Figure~\ref{fig:fitting3dchic1} shows the fit result of \CHIC. Figure~\ref{fig:fitting3d4} shows the fit result of \PLPHI. After applying charmonia veto, the fit results of \LLKO \ and \LLKS \ are shown in Fig.~\ref{fig:fitting3dL}.

\begin{figure}[htbp]
\centering
{
\includegraphics[width=4.2cm]{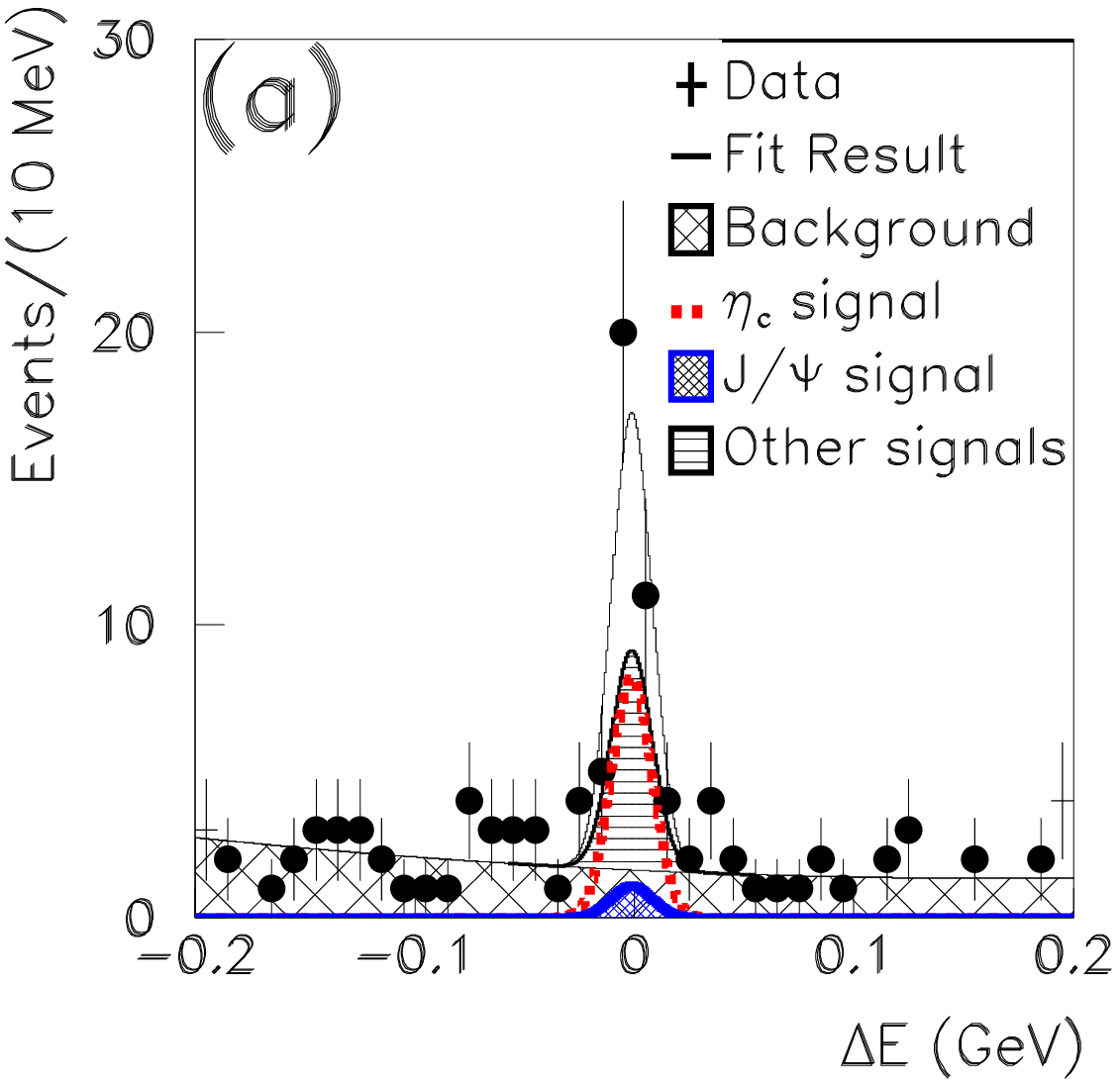}
\includegraphics[width=4.2cm]{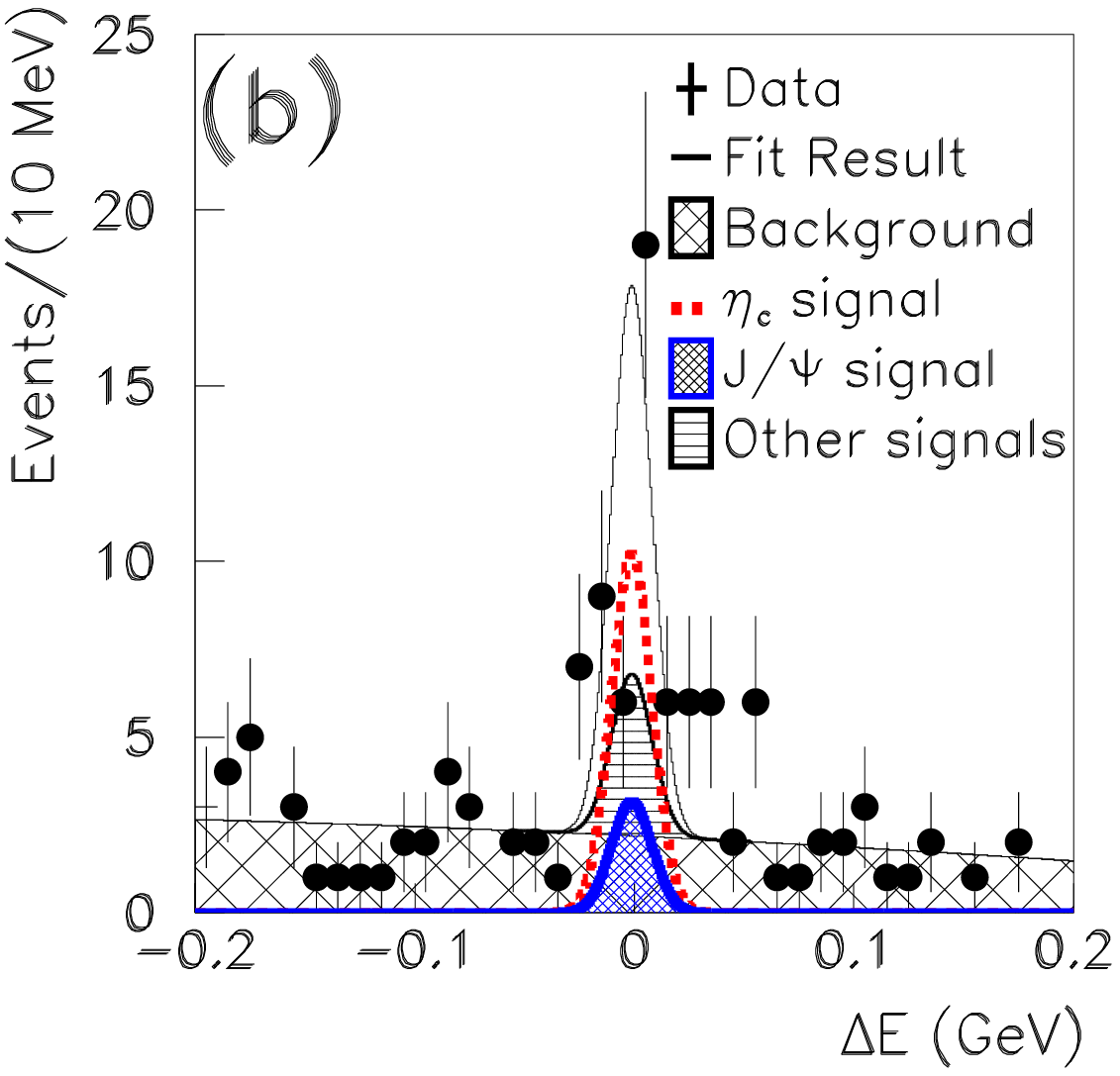}
\includegraphics[width=4.2cm]{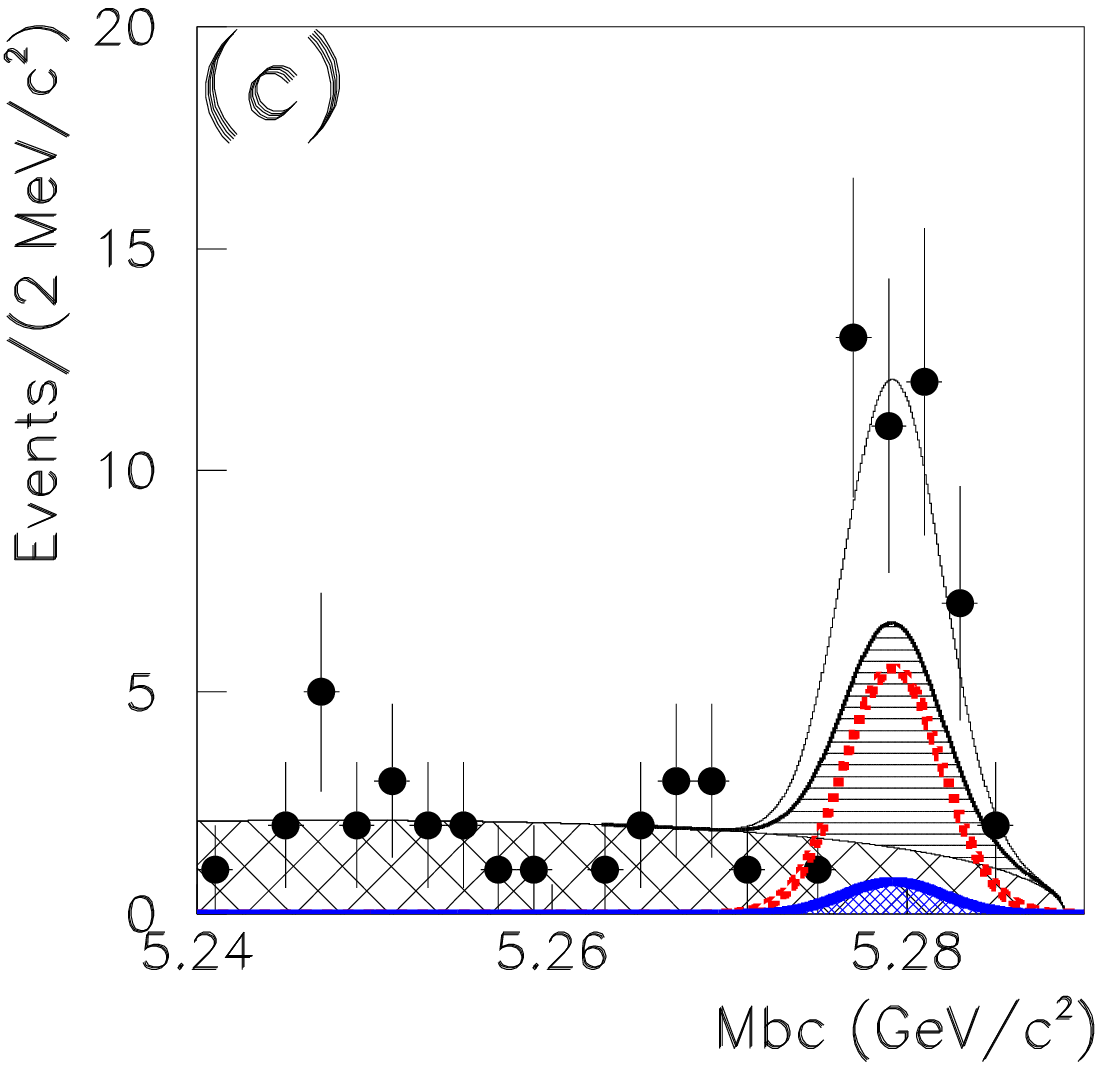}
\includegraphics[width=4.2cm]{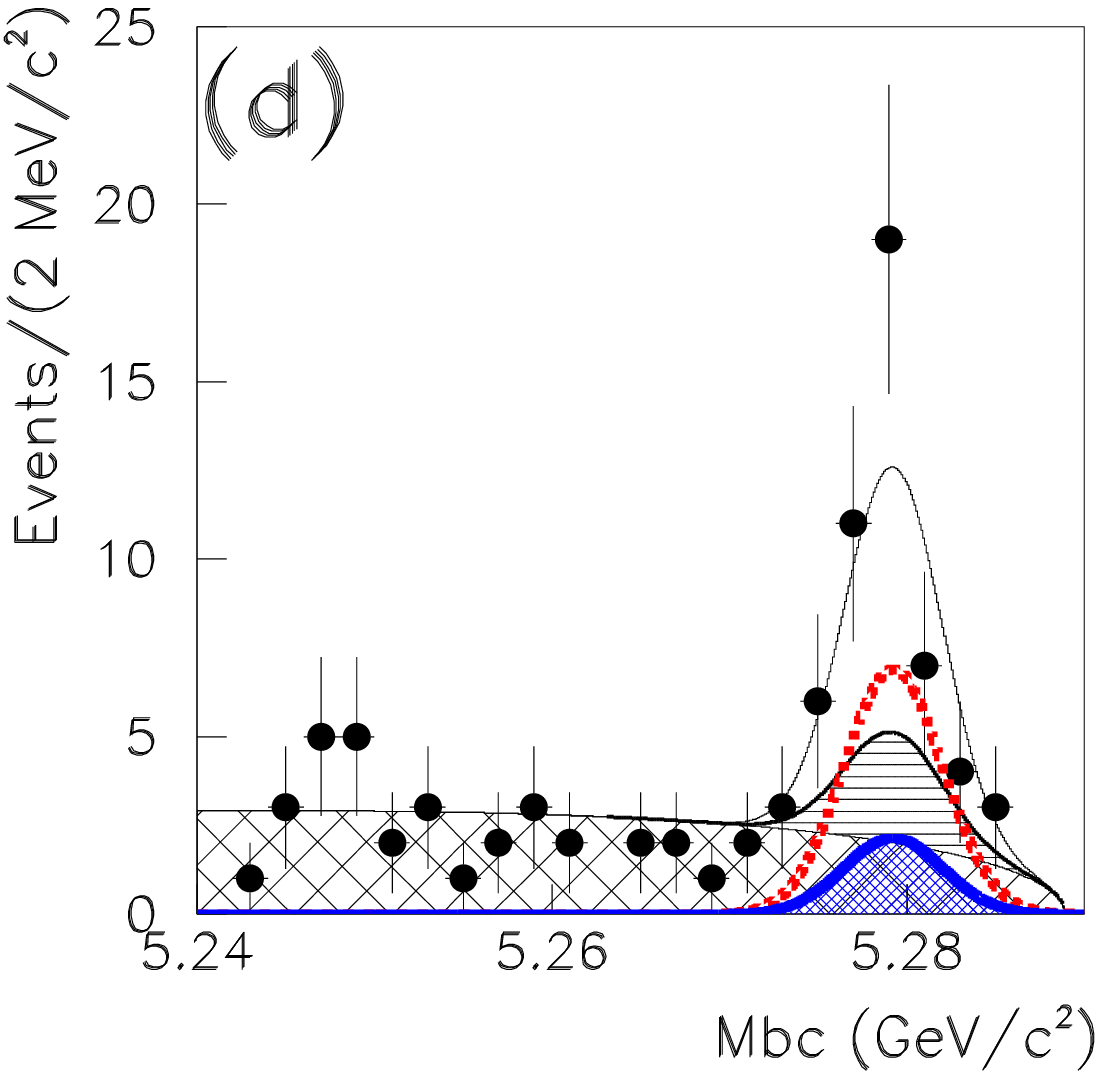}
\includegraphics[width=4.2cm]{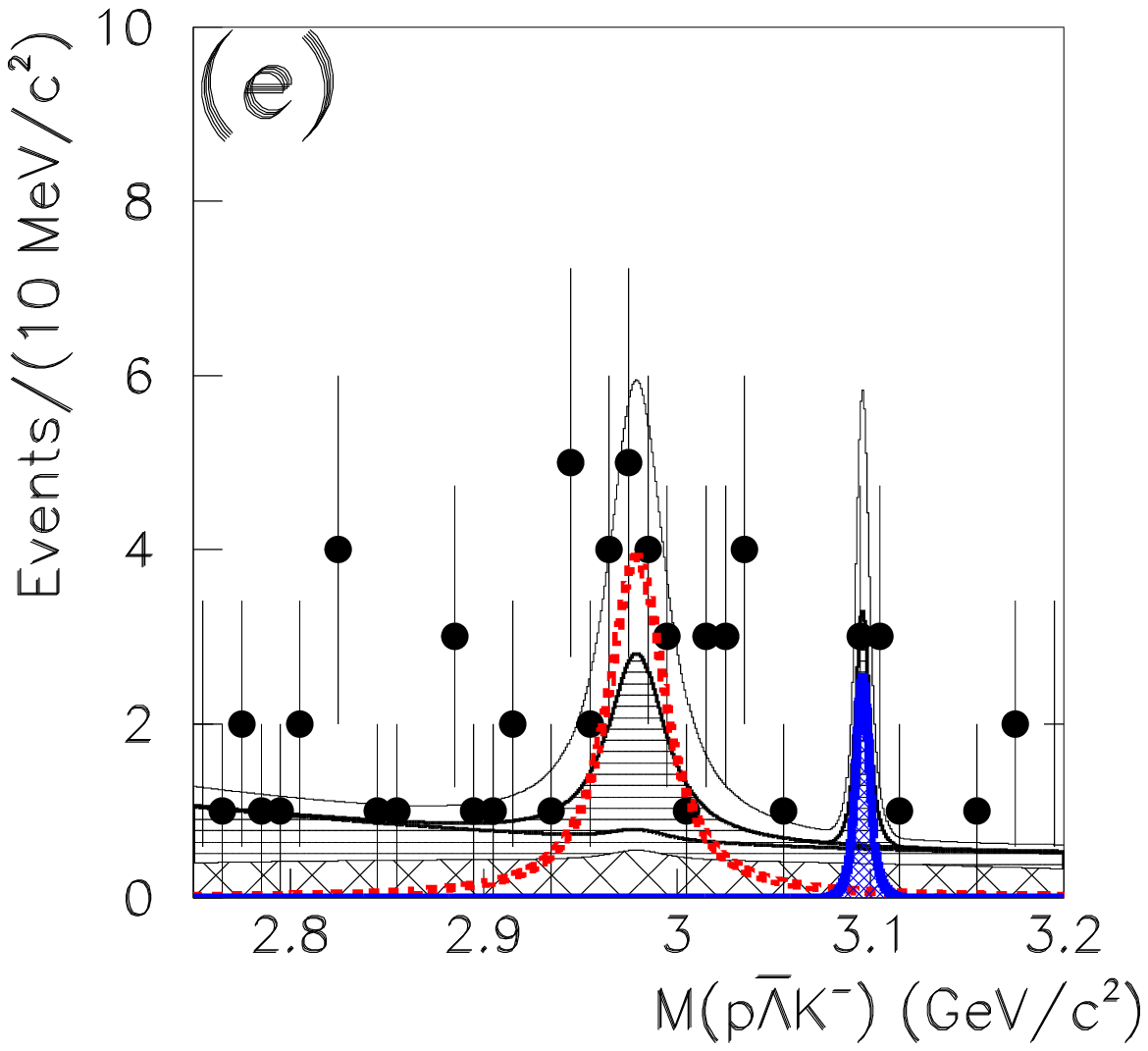}
\includegraphics[width=4.2cm]{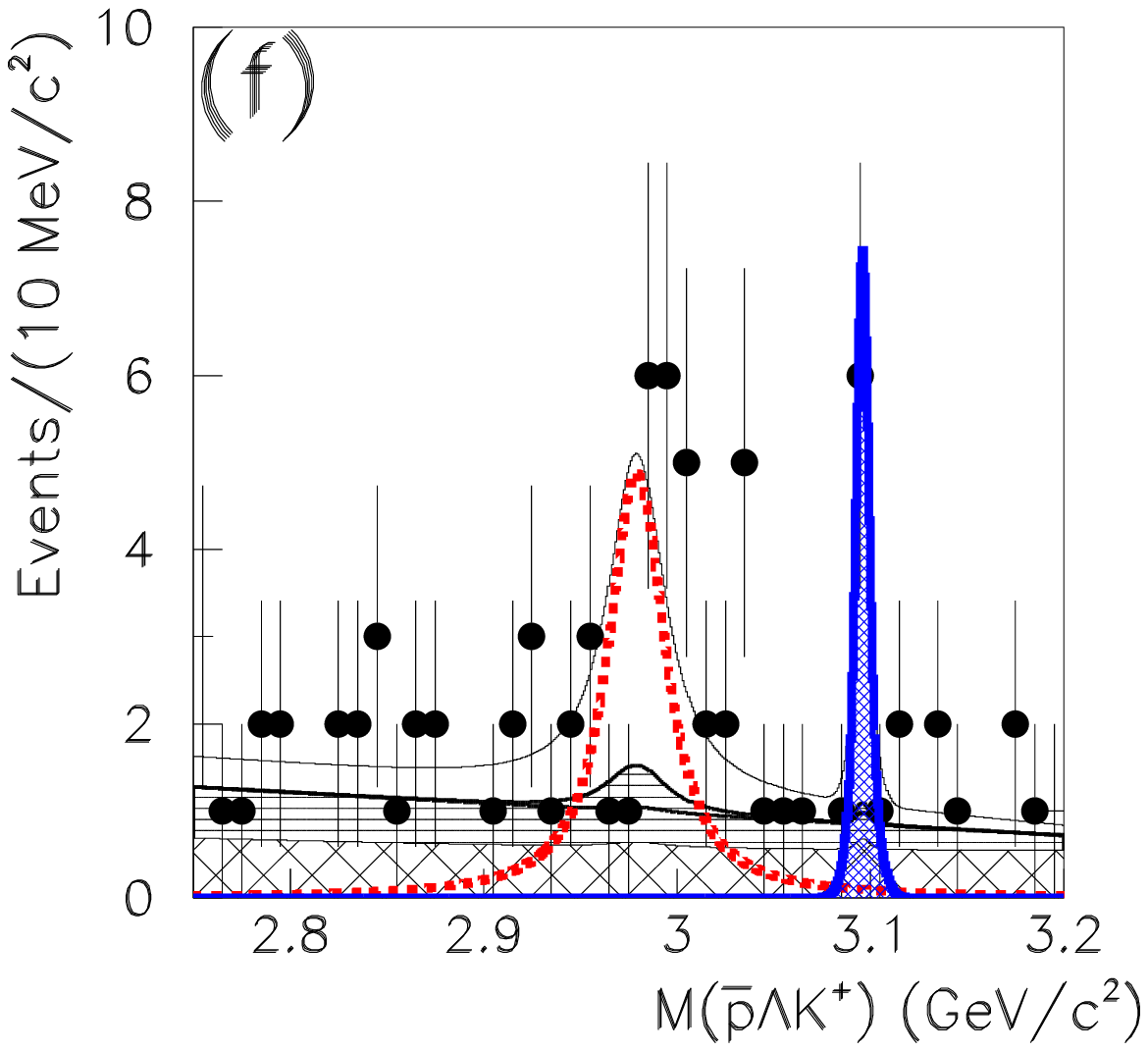}
\includegraphics[width=4.2cm]{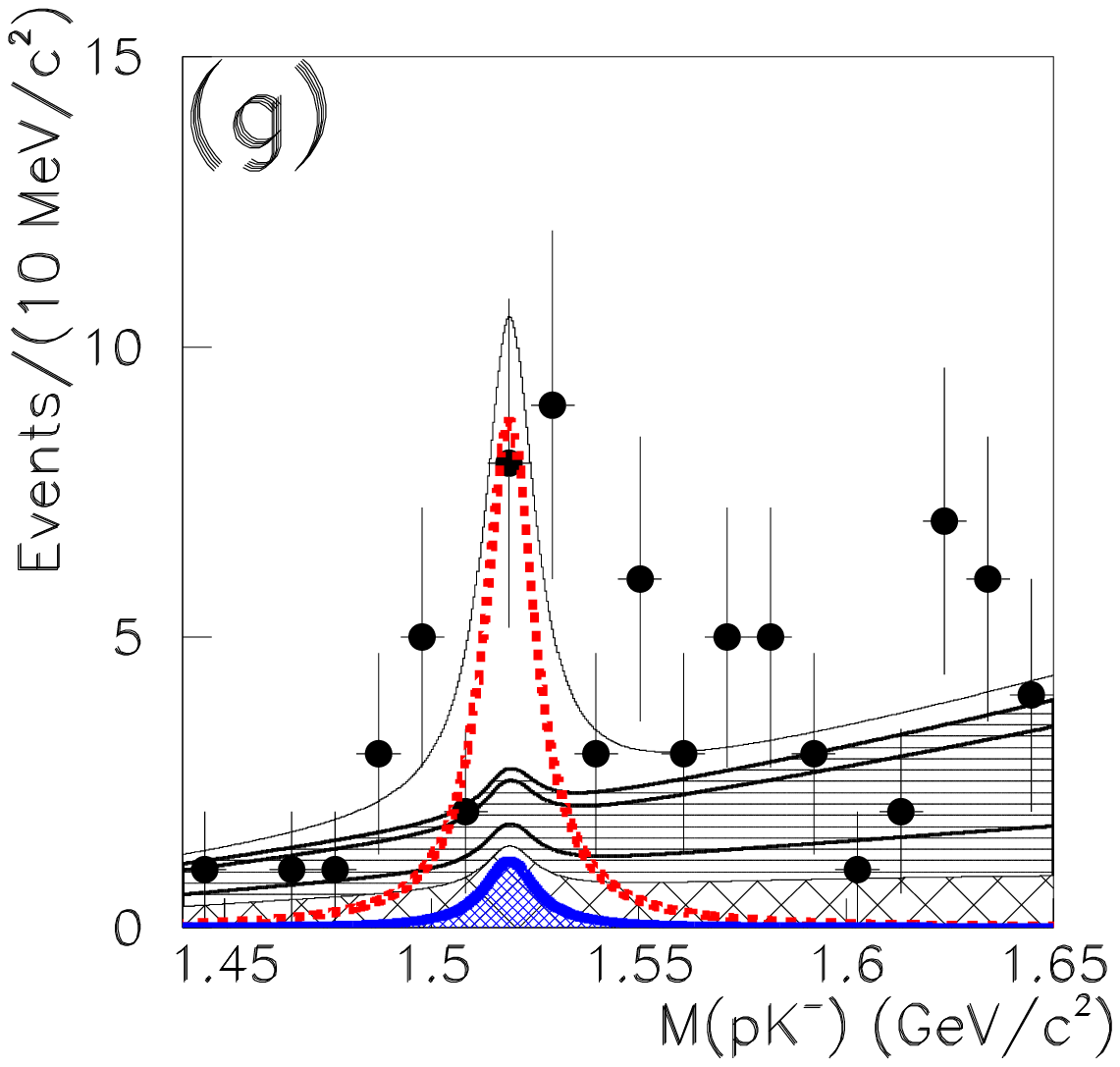}
\includegraphics[width=4.2cm]{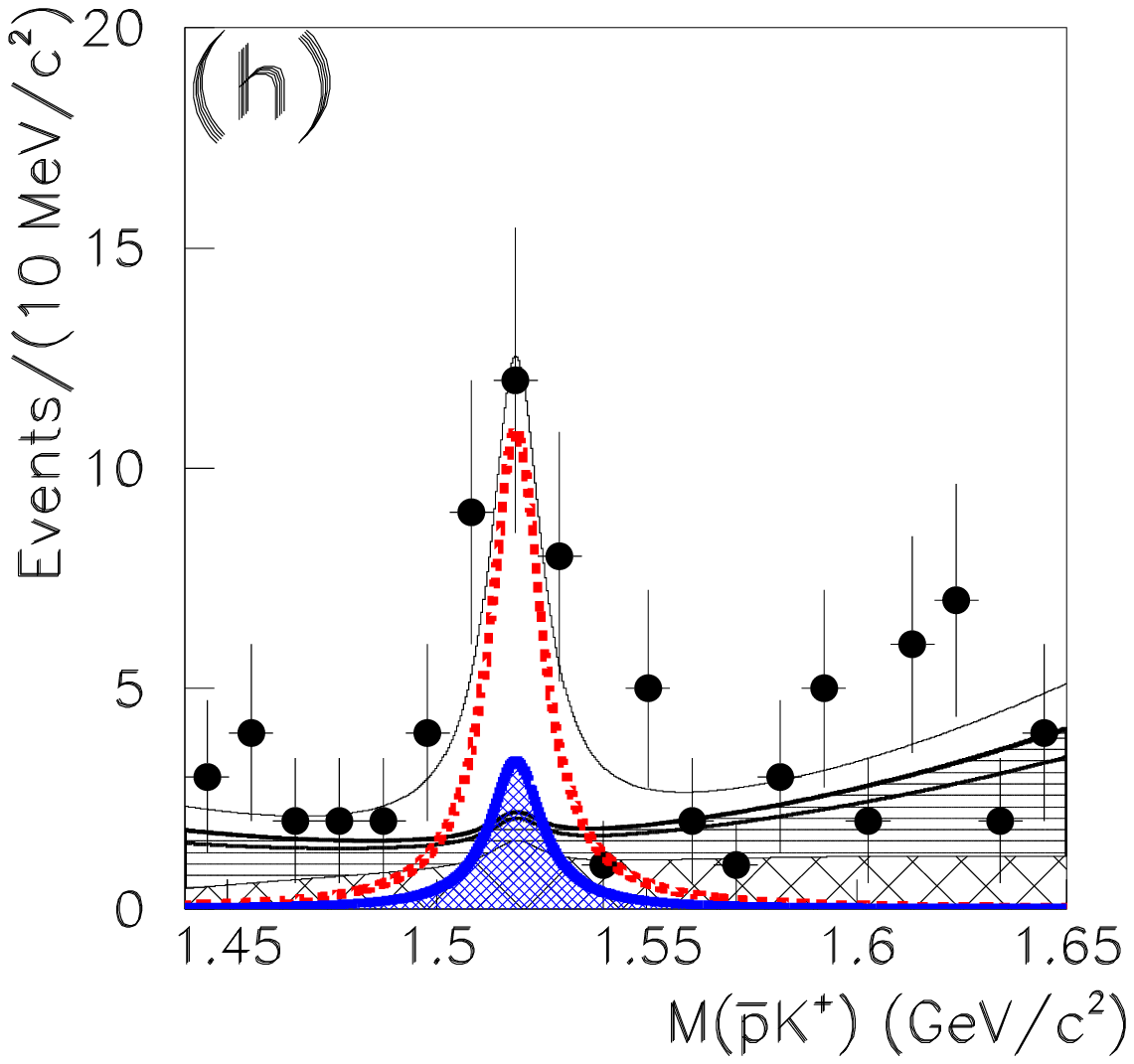}
}
\caption{Fit results of  \ETAC \ (\ETACS) and \JPSI \ (\JPSIS) in projection plots of $\Delta E$ (5.27 $<M_{bc}<$ 5.29, 2.9 $<M_{p\bar{\Lambda}K^-}/M_{\bar{p} \Lambda K^+}<$ 3.12 \GeV \ and 1.45 $<M_{pK^-}/M_{\bar{p}K^+}<$ 1.58 \GeV), $M_{bc}$ ($|\Delta E|<$ 0.03 GeV, 2.9 $<M_{p\bar{\Lambda}K^-}/M_{\bar{p} \Lambda K^+}<$ 3.12 \GeV \ and 1.45 $<M_{pK^-}/M_{\bar{p}K^+}<$ 1.58 \GeV), $M_{p\bar{\Lambda}K^-}/M_{\bar{p} \Lambda K^+}$ (in signal box and 1.45 $<M_{pK^-}/M_{\bar{p}K^+}<$ 1.58 \GeV) and $M_{pK^-}/M_{\bar{p}K^+}$ (in signal box and 2.9 $<M_{p\bar{\Lambda}K^-}/M_{\bar{p} \Lambda K^+}<$ 3.12 \GeV). (a)(c)(e)(g) are for the final state $p \bar{\Lambda} K^+ K^-$; (b)(d)(f)(h) are for the final state $\bar{p} \Lambda K^+ K^+$. For illustration purpose, we only show signal curve peaking in all spectra,  and merge other $B$ decay signals as horizontal-line region and all backgrounds as cross-hatched region.}
\label{fig:fitting4d}
\end{figure}

In the mass window of $\eta_c$, we observe a clear resonance in $M(pK^-)$, at the nominal mass of $\Lambda(1520)$, indicating a non-negligible fraction of \ETACD \ from \ETACS. In the same manner, we fit the $\Delta E$, $M_{bc}$, $M(p\bar{\Lambda}K^-)$ and $M(pK^-)$ spectra simultaneously in order to determine the yields of \ETACS \ and \JPSIS. The fit results are shown in Fig.~\ref{fig:fitting4d}.

The value of the signal significance is defined by $\sqrt{-2\times\ln({\mathcal{L}_0}/{\mathcal{L}_s})} (\sigma)$, where $\mathcal{L}_0$ is the likelihood with null signal yield and  $\mathcal{L}_s$ is the likelihood with measured yield. In the above calculation, we have used the likelihood function which is smeared by considering the additive systematic uncertainties that would affect the fitted yield. For those modes with signal significance less than $3\sigma$, we integrate the smeared likelihood function in order to find out the upper limit yield at the 90\% confidence level. That is, to calculate N that satisfies

$$\int_{0}^{N}\mathcal{L}(n)dn=0.9\int_{0}^{\infty}\mathcal{L}(n)dn,$$
where $\mathcal{L}(n)$ denotes the likelihood function with the condition that the number of signal events is fixed to the value $n$.

For systematic uncertainty, we consider tracking uncertainty per charged track (0.35\% per track and 0.70\% for $\Lambda$). The uncertainty of the estimated number of $B\bar{B}$ pairs is 1.4$\%$. The $\Lambda$ selection uncertainty is determined by the difference of the flight-distance distribution between data and MC (3.0\%). Some of systematic uncertainties are mode-dependent. The uncertainty in proton/antiproton identification is determined by using the study of $\Lambda$/$\bar{\Lambda}$ ($0.38\%$ to $0.53\%$) in data, while the uncertainty in kaon identification is determined from the study of $D^{*+} \rightarrow D^0\pi^+$, $D^0 \rightarrow K^- \pi^+$ in data ($2.0\%$ to $3.7\%$). We generate two kinds of signal MC: one considering a threshold enhancement in the dibaryonic system, the other with only phase space decays, and we mix the two samples to mimic the real data. The MC modeling uncertainty is set to be the larger difference in reconstruction efficiency between the threshold enhancement MC and phase space MC ($0.52\%$ to $9.3\%$). The smallest value, 0.52\%, is for \ETAC \ due to limited phase space. The uncertainty from the fixed signal probability density function is obtained by varying all of the shape variables by one sigma and refitting ($2.7\%$ to $3.3\%$). The statistical uncertainty of the MC reconstruction efficiency is $0.31\%$ to $0.47\%$. The uncertainty of $q\bar{q}$ suppression is obtained from the reconstruction efficiency difference with and without the cut ($0.50\%$ to $5.0\%$). We apply the $D^0$ veto to redo the analysis and attribute the possible veto uncertainty $2.2\%$ to $7.4\%$, where the statistical uncertainty from data is included. All the above uncertainties are combined in quadrature to obtain the total systematic uncertainties ($5.9\%$ to $12\%$).

Table~\ref{table:fitSummary} summarizes the fit yields, reconstruction efficiencies and corresponding systematic uncertainties of significant and evident modes; Table~\ref{table:fitSummary2} summarizes the upper limit yields and reconstruction efficiencies for modes with signal significance less than 3$\sigma$. Note that the reconstruction efficiencies in Table~\ref{table:fitSummary} and Table~\ref{table:fitSummary2} include the decay branching fraction 63.9\% for the long-lived $\Lambda \to p\pi^-$ in the MC simulation and efficiencies have been corrected for the MC-data difference of the proton/kaon identification.

We use the world average values~\cite{PDG16} of $\mathcal{B}(\Upsilon(4S) \to B^+B^-)$, $\mathcal{B}(\phi \to K^+K^-)$, $\mathcal{B}(\Lambda(1520) \to pK^-)$, $\mathcal{B}$(\ETAC), $\mathcal{B}$(\JPSI) and $\mathcal{B}$(\CHIC), to obtain the results listed in Table~\ref{table:Results}. The measured branching fractions of four-body decay of \OK \ and \PLPHI \ are consistent with theoretical predictions~\cite{HSIAO2017348,PhysRevD.85.017501}. Note that $\mathcal{B}$($B^+ \to p \bar{\Lambda} K^+ K^-$) is compatible with $\mathcal{B}$($B^+ \to p \bar{\Lambda} \pi^+ \pi^-$)~\cite{Chen:2009xg}.

\begin{table}[htbp]
\begin{center}
\renewcommand\arraystretch{1.4}
\scriptsize
\caption{Signal yields ($N_s$), reconstruction efficiencies ($\varepsilon_{\rm eff}$), systematic uncertainties ($\rm sys$) and significances (sig) from extended unbinned maximum likelihood fits for modes with signal significance greater than 3$\sigma$.}
\label{table:fitSummary}
\begin{tabular}[t]{l|ccc|c}
\hline\hline
Mode & $N_s$ &$\varepsilon_{\rm eff}$(\%)&$\rm sys$(\%)&sig ($\sigma$)\\
\hline
$B^+ \rightarrow p\bar{\Lambda} K^+ K^-$  & $190.1^{+20.3}_{-19.6}$ &5.84&12.2&11.7\\\hline
$B^+ \rightarrow  \bar{p}\Lambda  K^+ K^+$  & $188.0^{+19.2}_{-18.4}$ &6.40&11.8&12.7\\\hline\hline

($B^+\rightarrow\eta_c K^+$)  & $89.7^{+14.1}_{-13.3}$ &7.19&5.91&8.46\\
$\times$ ($\eta_c \rightarrow p\bar{\Lambda} K^-$)&&&&\\\hline
($B^+\rightarrow\eta_c K^+$)  & $67.0^{+14.1}_{-13.3}$&7.36&7.55&5.63\\
$\times$ ($\eta_c \rightarrow \bar{p}\Lambda K^+$)&&&&\\\hline
\multicolumn{4}{c|}{Total significance of the $\eta_c$ mode}&10.2\\\hline\hline

($B^+\rightarrow J/\psi K^+$)  &$19.0^{+5.7}_{-5.0}$ &6.57&7.83&4.92\\
$\times$ ($J/\psi \rightarrow p\bar{\Lambda} K^-$)&&&&\\\hline
($B^+\rightarrow J/\psi K^+$)  &$25.5^{+6.6}_{-5.9}$ &6.56&5.90&5.50\\
$\times$ ($J/\psi \rightarrow \bar{p}\Lambda K^+$)&&&&\\\hline
\multicolumn{4}{c|}{Total significance of the $J/\psi$ mode}&7.38\\\hline\hline

($B^+\rightarrow\chi_{c1} K^+$)  & $10.2^{+4.6}_{-3.9}$ &7.39&11.9&3.18\\
$\times$ ($\chi_{c1} \rightarrow p\bar{\Lambda} K^-$)&&&&\\\hline
($B^+\rightarrow\chi_{c1} K^+$)  & $13.4^{+5.0}_{-4.3}$&6.38&10.5&3.79\\
$\times$ ($\chi_{c1} \rightarrow \bar{p}\Lambda K^+$)&&&&\\\hline
\multicolumn{4}{c|}{Total significance of the $\chi_{c1}$ mode}&4.95\\\hline\hline

($B^+ \rightarrow  p\bar{\Lambda}\phi$) & 23.2$\pm$6.1&7.52&9.53&5.15\\
$\times$ ($\phi \rightarrow K^+K^-$)&&&&\\\hline\hline

($B^+ \rightarrow  \Lambda(1520)\bar{\Lambda} K^+$) & 30.3$\pm$8.6&7.60&10.5&4.08\\
$\times$ ($\Lambda(1520) \rightarrow p K^-$)&&&&\\\hline\hline

($B^+\rightarrow\eta_c K^+$)  & $19.2\pm 12.5$ &7.58&9.68&1.97\\
$\times$ ($\eta_c \rightarrow \Lambda(1520)\bar{\Lambda}$)&&&\\
$\times$ ($\Lambda(1520) \rightarrow p K^-$)&&&\\\hline
($B^+\rightarrow\eta_c K^+$)  & $23.9\pm 13.4$&6.95&6.40&2.50\\
$\times$ ($\eta_c \rightarrow \bar{\Lambda}(1520)\Lambda$)&&&\\
$\times$ ($\bar{\Lambda}(1520) \rightarrow \bar{p}K^+$)&&&\\\hline
\multicolumn{4}{c|}{Total significance of the $\eta_c$ sub-mode}&3.18\\

\hline\hline
\end{tabular}
\end{center}
\end{table}

\begin{table}[htbp]
\begin{center}
\renewcommand\arraystretch{1.4}
\scriptsize
\caption{Upper limits of yields ($N_{upper}$) and reconstruction efficiencies ($\varepsilon_{\rm eff}$) from extended unbinned maximum likelihood fits for modes with signal significance less than 3$\sigma$. For the $J/\psi$ decay, we determine its upper limit of branching fraction with the combined \OK \ and \SK \ data samples.}
\label{table:fitSummary2}
\begin{tabular}[t]{l|cc|c}
\hline\hline
Mode & $N_{upper}$ &$\varepsilon_{eff}$(\%)&comment\\\hline
($B^+\rightarrow J/\psi K^+$)  & 17.2&5.88&90\% C.L.\\
$\times$ ($J/\psi \rightarrow \Lambda(1520)\bar{\Lambda}$)&&&\\
$\times$ ($\Lambda(1520) \rightarrow p K^-$)&&&\\\hline\hline
($B^+\rightarrow\bar{\Lambda}(1520)\Lambda K^+$)& 19.8&5.70&90\% C.L.\\
$\times$ ($\bar{\Lambda}(1520) \rightarrow \bar{p}K^+$)&&&\\
\hline\hline
\end{tabular}
\end{center}
\end{table}

%Points with error bars are data, the cross-hatched region covering the whole bottom is background, the dotted/darker-hatched line on the bottom is $\eta_c$/$J/\psi$ signals,  the horizontal-line region accumulating on background is the contribution of four-body decay, the solid black curve is the total distribution of all components. In this fit, the feed-down of background of resonances is negligible.

%Points with error bars are data, the dotted line is signal, the cross-hatched region covering the whole bottom is background, the horizontal-line region accumulating on background is contributions of four-body decay, the solid black curve is the total distribution of all components.

%Points with error bars are data, the cross-hatched region covering the whole bottom is background, the dotted/darker-hatched line on the bottom is $\Lambda(1520)$ signal,  the horizontal-line region accumulating on background is the contribution of four-body decay, other $\eta_c$ and J$\psi$ decays, the solid black curve is the total distribution of all components.

%2.9 $<M_{p\bar{\Lambda}K^-}<$ 3.12 \GeV and 1.45 $<M_{pK^-}<$ 1.58 \GeV

\begin{table}[htbp]
\begin{center}
\renewcommand\arraystretch{1.4}
\footnotesize
\caption{Summary of measured branching fractions. Here c.c. stands for the corresponding charge-conjugation process. The listed four-body modes exclude any intermediate resonance.}
\label{table:Results}
\begin{tabular}[t]{l|c}
\hline\hline
Mode&Branching fraction\\\hline
$B^+ \rightarrow p\bar{\Lambda} K^+ K^-$& $(4.10^{+0.45}_{-0.43}\pm 0.50)\times10^{-6}$\\\hline
$B^+ \rightarrow \bar{p}\Lambda K^+ K^+$& $(3.70^{+0.39}_{-0.37}\pm 0.44)\times10^{-6}$\\\hline
$B^+ \rightarrow  p\bar{\Lambda}\phi$&$(7.95 \pm 2.09 \pm 0.77)\times10^{-7}$\\\hline
$\eta_c \rightarrow p\bar{\Lambda} K^-$ + c.c.&$(2.83^{+0.36}_{-0.34}\pm 0.35)\times10^{-3}$\\\hline
$J/\psi \rightarrow p\bar{\Lambda} K^-$ + c.c.&$(8.32^{+1.63}_{-1.45}\pm 0.49)\times10^{-4}$\\\hline
$\chi_{c1}\rightarrow p\bar{\Lambda} K^-$ + c.c.&$(9.15^{+2.63}_{-2.25}\pm 0.86)\times10^{-4}$\\\hline
$B^+\rightarrow\Lambda(1520)\bar{\Lambda} K^+$&$(2.23 \pm 0.63 \pm 0.25)\times 10^{-6}$\\\hline
$\eta_c \rightarrow \Lambda(1520)\bar{\Lambda}$ + c.c.&$(3.48 \pm 1.48 \pm 0.46)\times 10^{-3}$\\\hline
$J/\psi \rightarrow \Lambda(1520)\bar{\Lambda}$ + c.c.&$<1.80\times10^{-3}$\\\hline
$B^+\rightarrow\bar{\Lambda}(1520)\Lambda K^+$&$<2.08\times10^{-6}$\\
\hline\hline
\end{tabular}
\end{center}
\end{table}

In summary, using a sample of $772 \times 10^6$ $B\bar{B}$ pair events, we measure the branching fractions of the four-body decays \OK \ and \SK \  with intermediate resonance modes being excluded. The feature of a threshold enhancement of the dibaryon system persists, but with a non-negligible phase space contribution. We also observe the three-body decay of \ETACD+c.c. The measured $\mathcal{B}$(\JPSID+c.c.) is in good agreement with the world average~\cite{PDG16}. We also confirm the observation of \CHICD+c.c. These decay amplitudes can be useful for a better understanding of the charmonium system. We observe the charmless decay \PLPHI \ with a smaller branching fraction than that of the four-body decay. Its signal yield is not significant enough to perform an angular analysis.

We thank the KEKB group for excellent operation of the
accelerator; the KEK cryogenics group for efficient solenoid
operations; and the KEK computer group, the NII, and
PNNL/EMSL for valuable computing and SINET5 network support.
We acknowledge support from MEXT, JSPS and Nagoya's TLPRC (Japan);
ARC (Australia); FWF (Austria); NSFC and CCEPP (China);
MSMT (Czechia); CZF, DFG, EXC153, and VS (Germany);
DST (India); INFN (Italy);
MOE, MSIP, NRF, RSRI, FLRFAS project and GSDC of KISTI (Korea);
MNiSW and NCN (Poland); MES under contract 14.W03.31.0026 (Russia); ARRS (Slovenia);
IKERBASQUE and MINECO (Spain);
SNSF (Switzerland); MOE and MOST (Taiwan); and DOE and NSF (USA).

\bibliography{refpage}

\begin{thebibliography}{27}%
\makeatletter
\providecommand \@ifxundefined [1]{%
 \@ifx{#1\undefined}
}%
\providecommand \@ifnum [1]{%
 \ifnum #1\expandafter \@firstoftwo
 \else \expandafter \@secondoftwo
 \fi
}%
\providecommand \@ifx [1]{%
 \ifx #1\expandafter \@firstoftwo
 \else \expandafter \@secondoftwo
 \fi
}%
\providecommand \natexlab [1]{#1}%
\providecommand \enquote  [1]{``#1''}%
\providecommand \bibnamefont  [1]{#1}%
\providecommand \bibfnamefont [1]{#1}%
\providecommand \citenamefont [1]{#1}%
\providecommand \href@noop [0]{\@secondoftwo}%
\providecommand \href [0]{\begingroup \@sanitize@url \@href}%
\providecommand \@href[1]{\@@startlink{#1}\@@href}%
\providecommand \@@href[1]{\endgroup#1\@@endlink}%
\providecommand \@sanitize@url [0]{\catcode `\\12\catcode `\$12\catcode
  `\&12\catcode `\#12\catcode `\^12\catcode `\_12\catcode `\%12\relax}%
\providecommand \@@startlink[1]{}%
\providecommand \@@endlink[0]{}%
\providecommand \url  [0]{\begingroup\@sanitize@url \@url }%
\providecommand \@url [1]{\endgroup\@href {#1}{\urlprefix }}%
\providecommand \urlprefix  [0]{URL }%
\providecommand \Eprint [0]{\href }%
\providecommand \doibase [0]{http://dx.doi.org/}%
\providecommand \selectlanguage [0]{\@gobble}%
\providecommand \bibinfo  [0]{\@secondoftwo}%
\providecommand \bibfield  [0]{\@secondoftwo}%
\providecommand \translation [1]{[#1]}%
\providecommand \BibitemOpen [0]{}%
\providecommand \bibitemStop [0]{}%
\providecommand \bibitemNoStop [0]{.\EOS\space}%
\providecommand \EOS [0]{\spacefactor3000\relax}%
\providecommand \BibitemShut  [1]{\csname bibitem#1\endcsname}%
\let\auto@bib@innerbib\@empty
%</preamble>
\bibitem [{\citenamefont {Bevan}\ \emph {et~al.}(2015)\citenamefont {Bevan}
  \emph {et~al.}}]{Bevan2014}%
  \BibitemOpen
  \bibfield  {author} {\bibinfo {author} {\bibfnamefont {A.~J.}\ \bibnamefont
  {Bevan}} \emph {et~al.},\ }\href@noop {} {\emph {\bibinfo {title} {The
  Physics of the B Factories}}},\ \bibinfo {edition} {4th}\ ed.\ (\bibinfo
  {publisher} {Springer-Verlag},\ \bibinfo {year} {2015})\BibitemShut {NoStop}%
\bibitem [{\citenamefont {Patrignani}\ \emph {et~al.}(2016)\citenamefont
  {Patrignani} \emph {et~al.}}]{PDG16}%
  \BibitemOpen
  \bibfield  {author} {\bibinfo {author} {\bibfnamefont {C.}~\bibnamefont
  {Patrignani}} \emph {et~al.} (\bibinfo {collaboration} {Particle Data
  Group}),\ }\href@noop {} {\bibfield  {journal} {\bibinfo  {journal} {Chin.
  Phys. C}\ }\textbf {\bibinfo {volume} {40}},\ \bibinfo {pages} {100001}
  (\bibinfo {year} {2016})}\BibitemShut {NoStop}%
%%CITATION = HEP-EX/0701057;%%
\bibitem [{Note1()}]{Note1}%
  \BibitemOpen
  \bibinfo {note} {Through out this paper, inclusion of charge-conjugate mode
  is always implied if the charge-conjugate final states are not specifically
  mentioned together.}\BibitemShut {Stop}%
\bibitem [{\citenamefont {Aaij}\ \emph
  {et~al.}(2017{\natexlab{a}})\citenamefont {Aaij} \emph
  {et~al.}}]{PhysRevLett.119.232001}%
  \BibitemOpen
  \bibfield  {author} {\bibinfo {author} {\bibfnamefont {R.}~\bibnamefont
  {Aaij}} \emph {et~al.} (\bibinfo {collaboration} {LHCb Collaboration}),\
  }\href {\doibase 10.1103/PhysRevLett.119.232001} {\bibfield  {journal}
  {\bibinfo  {journal} {Phys. Rev. Lett.}\ }\textbf {\bibinfo {volume} {119}},\
  \bibinfo {pages} {232001} (\bibinfo {year} {2017}{\natexlab{a}})}\BibitemShut
  {NoStop}%
\bibitem [{\citenamefont {Aaij}\ \emph
  {et~al.}(2014{\natexlab{a}})\citenamefont {Aaij} \emph
  {et~al.}}]{Aaij:2014tua}%
  \BibitemOpen
  \bibfield  {author} {\bibinfo {author} {\bibfnamefont {R.}~\bibnamefont
  {Aaij}} \emph {et~al.} (\bibinfo {collaboration} {LHCb Collaboration}),\
  }\href {\doibase 10.1103/PhysRevLett.113.141801} {\bibfield  {journal}
  {\bibinfo  {journal} {Phys. Rev. Lett.}\ }\textbf {\bibinfo {volume} {113}},\
  \bibinfo {pages} {141801} (\bibinfo {year} {2014}{\natexlab{a}})}\BibitemShut
  {NoStop}%
%%CITATION = ARXIV:1407.5907;%%
\bibitem [{\citenamefont {Aaij}\ \emph
  {et~al.}(2017{\natexlab{b}})\citenamefont {Aaij} \emph
  {et~al.}}]{PhysRevLett.119.041802}%
  \BibitemOpen
  \bibfield  {author} {\bibinfo {author} {\bibfnamefont {R.}~\bibnamefont
  {Aaij}} \emph {et~al.} (\bibinfo {collaboration} {LHCb Collaboration}),\
  }\href {\doibase 10.1103/PhysRevLett.119.041802} {\bibfield  {journal}
  {\bibinfo  {journal} {Phys. Rev. Lett.}\ }\textbf {\bibinfo {volume} {119}},\
  \bibinfo {pages} {041802} (\bibinfo {year} {2017}{\natexlab{b}})}\BibitemShut
  {NoStop}%
\bibitem [{\citenamefont {Aaij}\ \emph
  {et~al.}(2014{\natexlab{b}})\citenamefont {Aaij} \emph
  {et~al.}}]{PhysRevLett.113.152003}%
  \BibitemOpen
  \bibfield  {author} {\bibinfo {author} {\bibfnamefont {R.}~\bibnamefont
  {Aaij}} \emph {et~al.} (\bibinfo {collaboration} {LHCb Collaboration}),\
  }\href {\doibase 10.1103/PhysRevLett.113.152003} {\bibfield  {journal}
  {\bibinfo  {journal} {Phys. Rev. Lett.}\ }\textbf {\bibinfo {volume} {113}},\
  \bibinfo {pages} {152003} (\bibinfo {year} {2014}{\natexlab{b}})}\BibitemShut
  {NoStop}%
\bibitem [{\citenamefont {Aaij}\ \emph
  {et~al.}(2017{\natexlab{c}})\citenamefont {Aaij} \emph
  {et~al.}}]{PhysRevD.96.051103}%
  \BibitemOpen
  \bibfield  {author} {\bibinfo {author} {\bibfnamefont {R.}~\bibnamefont
  {Aaij}} \emph {et~al.} (\bibinfo {collaboration} {LHCb Collaboration}),\
  }\href {\doibase 10.1103/PhysRevD.96.051103} {\bibfield  {journal} {\bibinfo
  {journal} {Phys. Rev. D}\ }\textbf {\bibinfo {volume} {96}},\ \bibinfo
  {pages} {051103} (\bibinfo {year} {2017}{\natexlab{c}})}\BibitemShut
  {NoStop}%
\bibitem [{\citenamefont {Chen}\ \emph {et~al.}(2008)\citenamefont {Chen},
  \citenamefont {Cheng}, \citenamefont {Geng},\ and\ \citenamefont
  {Hsiao}}]{C.Chen08}%
  \BibitemOpen
  \bibfield  {author} {\bibinfo {author} {\bibfnamefont {C.-H.}\ \bibnamefont
  {Chen}}, \bibinfo {author} {\bibfnamefont {H.-Y.}\ \bibnamefont {Cheng}},
  \bibinfo {author} {\bibfnamefont {C.~Q.}\ \bibnamefont {Geng}}, \ and\
  \bibinfo {author} {\bibfnamefont {Y.~K.}\ \bibnamefont {Hsiao}},\ }\href
  {\doibase 10.1103/PhysRevD.78.054016} {\bibfield  {journal} {\bibinfo
  {journal} {Phys. Rev. D}\ }\textbf {\bibinfo {volume} {78}},\ \bibinfo
  {pages} {054016} (\bibinfo {year} {2008})}\BibitemShut {NoStop}%
\bibitem [{\citenamefont {Y.-Y.Chang}\ \emph {et~al.}(2015)\citenamefont
  {Y.-Y.Chang} \emph {et~al.}}]{Y.-Y.Chang15}%
  \BibitemOpen
  \bibfield  {author} {\bibinfo {author} {\bibnamefont {Y.-Y.Chang}} \emph
  {et~al.} (\bibinfo {collaboration} {Belle Collaboration}),\ }\href {\doibase
  10.1103/PhysRevLett.115.221803} {\bibfield  {journal} {\bibinfo  {journal}
  {Phys. Rev. Lett.}\ }\textbf {\bibinfo {volume} {115}},\ \bibinfo {pages}
  {221803} (\bibinfo {year} {2015})}\BibitemShut {NoStop}%
\bibitem [{\citenamefont {Hsiao}\ and\ \citenamefont
  {Geng}(2016)}]{PhysRevD.93.034036}%
  \BibitemOpen
  \bibfield  {author} {\bibinfo {author} {\bibfnamefont {Y.~K.}\ \bibnamefont
  {Hsiao}}\ and\ \bibinfo {author} {\bibfnamefont {C.~Q.}\ \bibnamefont
  {Geng}},\ }\href {\doibase 10.1103/PhysRevD.93.034036} {\bibfield  {journal}
  {\bibinfo  {journal} {Phys. Rev. D}\ }\textbf {\bibinfo {volume} {93}},\
  \bibinfo {pages} {034036} (\bibinfo {year} {2016})}\BibitemShut {NoStop}%
\bibitem [{\citenamefont {Hsiao}\ and\ \citenamefont
  {Geng}(2017)}]{HSIAO2017348}%
  \BibitemOpen
  \bibfield  {author} {\bibinfo {author} {\bibfnamefont {Y.}~\bibnamefont
  {Hsiao}}\ and\ \bibinfo {author} {\bibfnamefont {C.}~\bibnamefont {Geng}},\
  }\href {\doibase https://doi.org/10.1016/j.physletb.2017.04.067} {\bibfield
  {journal} {\bibinfo  {journal} {Phys. Lett. B}\ }\textbf {\bibinfo {volume}
  {770}},\ \bibinfo {pages} {348 } (\bibinfo {year} {2017})}\BibitemShut
  {NoStop}%
\bibitem [{\citenamefont {Geng}\ and\ \citenamefont
  {Hsiao}(2012)}]{PhysRevD.85.017501}%
  \BibitemOpen
  \bibfield  {author} {\bibinfo {author} {\bibfnamefont {C.~Q.}\ \bibnamefont
  {Geng}}\ and\ \bibinfo {author} {\bibfnamefont {Y.~K.}\ \bibnamefont
  {Hsiao}},\ }\href {\doibase 10.1103/PhysRevD.85.017501} {\bibfield  {journal}
  {\bibinfo  {journal} {Phys. Rev. D}\ }\textbf {\bibinfo {volume} {85}},\
  \bibinfo {pages} {017501} (\bibinfo {year} {2012})}\BibitemShut {NoStop}%
\bibitem [{\citenamefont {Abashian}\ \emph {et~al.}(2002)\citenamefont
  {Abashian} \emph {et~al.}}]{Abashian02}%
  \BibitemOpen
  \bibfield  {author} {\bibinfo {author} {\bibfnamefont {A.}~\bibnamefont
  {Abashian}} \emph {et~al.} (\bibinfo {collaboration} {Belle Collaboration}),\
  }\href {\doibase https://doi.org/10.1016/S0168-9002(01)02013-7} {\bibfield
  {journal} {\bibinfo  {journal} {Nucl.\ Instrum.\ Methods Phys.\ Res.,\ Sect.\
  A}\ }\textbf {\bibinfo {volume} {479}},\ \bibinfo {pages} {117} (\bibinfo
  {year} {2002})}\BibitemShut {NoStop}%
\bibitem [{\citenamefont {Brodzicka}\ \emph {et~al.}(2012)\citenamefont
  {Brodzicka} \emph {et~al.}}]{Brodzicka12}%
  \BibitemOpen
  \bibfield  {author} {\bibinfo {author} {\bibfnamefont {J.}~\bibnamefont
  {Brodzicka}} \emph {et~al.},\ }\href@noop {} {\bibfield  {journal} {\bibinfo
  {journal} {Prog.\ Theor.\ Exp.\ Phys.}\ }\textbf {\bibinfo {volume} {04D001}}
  (\bibinfo {year} {2012})}\BibitemShut {NoStop}%
\bibitem [{\citenamefont {Kurokawa}\ and\ \citenamefont
  {Kikutani}(2003)}]{Kurokawa03}%
  \BibitemOpen
  \bibfield  {author} {\bibinfo {author} {\bibfnamefont {S.}~\bibnamefont
  {Kurokawa}}\ and\ \bibinfo {author} {\bibfnamefont {E.}~\bibnamefont
  {Kikutani}},\ }\href {\doibase https://doi.org/10.1016/S0168-9002(02)01771-0}
  {\bibfield  {journal} {\bibinfo  {journal} {Nucl.\ Instrum.\ Methods Phys.\
  Res.,\ Sect.\ A}\ }\textbf {\bibinfo {volume} {499}},\ \bibinfo {pages} {1 }
  (\bibinfo {year} {2003})}\BibitemShut {NoStop}%
\bibitem [{\citenamefont {Abe}\ \emph {et~al.}(2013)\citenamefont {Abe} \emph
  {et~al.}}]{Abe13}%
  \BibitemOpen
  \bibfield  {author} {\bibinfo {author} {\bibfnamefont {T.}~\bibnamefont
  {Abe}} \emph {et~al.},\ }\href@noop {} {\bibfield  {journal} {\bibinfo
  {journal} {Prog.\ Theor.\ Exp.\ Phys.}\ }\textbf {\bibinfo {volume} {03A001}}
  (\bibinfo {year} {2013})}\BibitemShut {NoStop}%
\bibitem [{\citenamefont {Nakano}\ \emph {et~al.}(2002)\citenamefont {Nakano}
  \emph {et~al.}}]{NAKANO2002402}%
  \BibitemOpen
  \bibfield  {author} {\bibinfo {author} {\bibfnamefont {E.}~\bibnamefont
  {Nakano}} \emph {et~al.} (\bibinfo {collaboration} {Belle Collaboration}),\
  }\href {\doibase https://doi.org/10.1016/S0168-9002(02)01510-3} {\bibfield
  {journal} {\bibinfo  {journal} {Nucl.\ Instrum.\ Methods Phys.\ Res.,\ Sect.\
  A}\ }\textbf {\bibinfo {volume} {494}},\ \bibinfo {pages} {402} (\bibinfo
  {year} {2002})}\BibitemShut {NoStop}%
\bibitem [{\citenamefont {Wei}\ \emph {et~al.}(2008)\citenamefont {Wei},
  \citenamefont {Wang}, \citenamefont {Adachi} \emph {et~al.}}]{ppk}%
  \BibitemOpen
  \bibfield  {author} {\bibinfo {author} {\bibfnamefont {J.-T.}\ \bibnamefont
  {Wei}}, \bibinfo {author} {\bibfnamefont {M.-Z.}\ \bibnamefont {Wang}},
  \bibinfo {author} {\bibfnamefont {I.}~\bibnamefont {Adachi}},  \emph {et~al.}
  (\bibinfo {collaboration} {Belle Collaboration}),\ }\href {\doibase
  https://doi.org/10.1016/j.physletb.2007.11.063} {\bibfield  {journal}
  {\bibinfo  {journal} {Phys. Lett. B}\ }\textbf {\bibinfo {volume} {659}},\
  \bibinfo {pages} {80 } (\bibinfo {year} {2008})}\BibitemShut {NoStop}%
\bibitem [{\citenamefont {Lange}(2001)}]{EvtGen}%
  \BibitemOpen
  \bibfield  {author} {\bibinfo {author} {\bibfnamefont {D.~J.}\ \bibnamefont
  {Lange}},\ }\href {\doibase https://doi.org/10.1016/S0168-9002(01)00089-4}
  {\bibfield  {journal} {\bibinfo  {journal} {Nucl.\ Instrum.\ Methods Phys.\
  Res.,\ Sect.\ A}\ }\textbf {\bibinfo {volume} {462}},\ \bibinfo {pages} {152
  } (\bibinfo {year} {2001})}\BibitemShut {NoStop}%
\bibitem [{\citenamefont {Brun}\ \emph {et~al.}(1984)\citenamefont {Brun} \emph
  {et~al.}}]{Geant}%
  \BibitemOpen
  \bibfield  {author} {\bibinfo {author} {\bibfnamefont {R.}~\bibnamefont
  {Brun}} \emph {et~al.},\ }\href@noop {} {\bibfield  {journal} {\bibinfo
  {journal} {CERN Report No. DD/EE/84-1}\ } (\bibinfo {year}
  {1984})}\BibitemShut {NoStop}%
%%CITATION = HEP-EX/0701057;%%
\bibitem [{\citenamefont {Feindt}\ and\ \citenamefont {Kerzel}(2006)}]{NB}%
  \BibitemOpen
  \bibfield  {author} {\bibinfo {author} {\bibfnamefont {M.}~\bibnamefont
  {Feindt}}\ and\ \bibinfo {author} {\bibfnamefont {U.}~\bibnamefont
  {Kerzel}},\ }\href {\doibase https://doi.org/10.1016/j.nima.2005.11.166}
  {\bibfield  {journal} {\bibinfo  {journal} {Nucl.\ Instrum.\ Methods Phys.\
  Res.,\ Sect.\ A}\ }\textbf {\bibinfo {volume} {559}},\ \bibinfo {pages} {190
  } (\bibinfo {year} {2006})}\BibitemShut {NoStop}%
\bibitem [{\citenamefont {Fox}\ and\ \citenamefont {S.Wolfram}(1978)}]{KSFW}%
  \BibitemOpen
  \bibfield  {author} {\bibinfo {author} {\bibfnamefont {G.}~\bibnamefont
  {Fox}}\ and\ \bibinfo {author} {\bibnamefont {S.Wolfram}},\ }\href {\doibase
  10.1103/PhysRevLett.41.1581} {\bibfield  {journal} {\bibinfo  {journal}
  {Phys. Rev. Lett.}\ }\textbf {\bibinfo {volume} {41}},\ \bibinfo {pages}
  {1581} (\bibinfo {year} {1978})}\BibitemShut {NoStop}%
\bibitem [{\citenamefont {Lee}\ \emph {et~al.}(2003)\citenamefont {Lee} \emph
  {et~al.}}]{KSFW2}%
  \BibitemOpen
  \bibfield  {author} {\bibinfo {author} {\bibfnamefont {S.~H.}\ \bibnamefont
  {Lee}} \emph {et~al.} (\bibinfo {collaboration} {Belle Collaboration}),\
  }\href@noop {} {\bibfield  {journal} {\bibinfo  {journal} {Phys. Rev Lett.}\
  }\textbf {\bibinfo {volume} {91}},\ \bibinfo {pages} {261801} (\bibinfo
  {year} {2003})}\BibitemShut {NoStop}%
%%CITATION = HEP-EX/0701057;%%
\bibitem [{\citenamefont {Kakuno}\ \emph {et~al.}(2004)\citenamefont {Kakuno}
  \emph {et~al.}}]{qr}%
  \BibitemOpen
  \bibfield  {author} {\bibinfo {author} {\bibfnamefont {H.}~\bibnamefont
  {Kakuno}} \emph {et~al.},\ }\href {\doibase
  https://doi.org/10.1016/j.nima.2004.06.159} {\bibfield  {journal} {\bibinfo
  {journal} {Nucl.\ Instrum.\ Methods Phys.\ Res.,\ Sect.\ A}\ }\textbf
  {\bibinfo {volume} {533}},\ \bibinfo {pages} {516 } (\bibinfo {year}
  {2004})}\BibitemShut {NoStop}%
\bibitem [{\citenamefont {Albrecht}\ \emph {et~al.}(1988)\citenamefont
  {Albrecht} \emph {et~al.}}]{argus}%
  \BibitemOpen
  \bibfield  {author} {\bibinfo {author} {\bibfnamefont {H.}~\bibnamefont
  {Albrecht}} \emph {et~al.} (\bibinfo {collaboration} {ARGUS Collaboration}),\
  }\href {\doibase https://doi.org/10.1016/0370-2693(88)90383-8} {\bibfield
  {journal} {\bibinfo  {journal} {Phys. Lett. B}\ }\textbf {\bibinfo {volume}
  {210}},\ \bibinfo {pages} {263 } (\bibinfo {year} {1988})}\BibitemShut
  {NoStop}%
\bibitem [{\citenamefont {Chen}\ \emph {et~al.}(2009)\citenamefont {Chen} \emph
  {et~al.}}]{Chen:2009xg}%
  \BibitemOpen
  \bibfield  {author} {\bibinfo {author} {\bibfnamefont {P.}~\bibnamefont
  {Chen}} \emph {et~al.} (\bibinfo {collaboration} {Belle}),\ }\href {\doibase
  10.1103/PhysRevD.80.111103} {\bibfield  {journal} {\bibinfo  {journal} {Phys.
  Rev. D}\ }\textbf {\bibinfo {volume} {80}},\ \bibinfo {pages} {111103}
  (\bibinfo {year} {2009})},\ \Eprint {http://arxiv.org/abs/0910.5817}
  {arXiv:0910.5817 [hep-ex]} \BibitemShut {NoStop}%
%%CITATION = ARXIV:0910.5817;%%
\end{thebibliography}%

\end{document}